\documentclass[fleqn,10pt]{wlscirep}
\usepackage[utf8]{inputenc}
\UseRawInputEncoding

\usepackage[T1]{fontenc}
\usepackage{soul}

\usepackage{xcolor}
\usepackage{tabularx}
\usepackage{epstopdf} 
\usepackage{epsfig}
\usepackage{textcomp}
\usepackage{xcolor}
\usepackage{subfigure}
\usepackage{graphicx}

\title{OPERAnet: A Multimodal Activity Recognition Dataset Acquired from Radio Frequency and Vision-based Sensors}

\author[1,*]{Mohammud J. Bocus}
\author[2,*]{Wenda Li}
\author[2,*]{Shelly Vishwakarma}
\author[1]{Roget Kou}
\author[2]{Chong Tang}
\author[2]{Karl Woodbridge}
\author[1]{Ian Craddock}
\author[1]{Ryan McConville}
\author[1]{Raul Santos-Rodriguez}
\author[2]{Kevin Chetty}
\author[1]{Robert Piechocki}

\affil[1]{School of Computer Science, Electrical and Electronic Engineering, and Engineering Maths, University of Bristol, Bristol, BS8 1UB, UK.}
\affil[2]{Department of Security and Crime Science, University College London, London 
WC1H 9EZ, UK}

\affil[*]{corresponding author(s): Mohammud J. Bocus (junaid.bocus@bristol.ac.uk), Wenda Li (wenda.li@ucl.ac.uk) and Shelly Vishwakarma (s.vishwakarma@ucl.ac.uk)}

\begin{abstract}

This paper presents a comprehensive dataset intended to evaluate passive Human Activity Recognition (HAR) and localization techniques with measurements obtained from synchronized Radio-Frequency (RF) devices and vision-based sensors. The dataset consists of RF data including Channel State Information (CSI) extracted from a WiFi Network Interface Card (NIC), Passive WiFi Radar (PWR) built upon a Software Defined Radio (SDR) platform, and Ultra-Wideband (UWB) signals acquired via commercial off-the-shelf hardware. It also consists of vision/Infra-red based data acquired from Kinect sensors. Approximately 8 hours of annotated measurements are provided, which are collected across two rooms from 6 participants performing 6 daily activities. This dataset can be exploited to advance WiFi and vision-based HAR, for example, using pattern recognition, skeletal representation, deep learning algorithms or other novel approaches to accurately recognize human activities. Furthermore, it can potentially be used to passively track a human in an indoor environment. Such datasets are key tools required for the development of new algorithms and methods in the context of smart homes, elderly care, and surveillance applications.

\end{abstract}

\begin{document}

\flushbottom
\maketitle

\thispagestyle{empty}


\section*{Background \& Summary}

Over the past few years, Human Activity Recognition (HAR) has become an active area of research due to its potential application in areas such as healthcare, Internet of Things (IoT), smart homes, surveillance, virtual gaming, among others. Numerous techniques have been proposed for HAR, ranging from wearable sensors such as accelerometers and/or gyroscopes \cite{accgyr}, vision-based methods such as Microsoft Xbox Kinect sensor \cite{ kinect2018}, to unobtrusive methods based on Radio-Frequency (RF) waves such as WiFi Channel State Information (CSI) \cite{translationresilientcsi}, Passive WiFi Radar (PWR) \cite{taxonomycsi}, Ultra-wideband (UWB) \cite{bocus2021uwb}, etc.  Recognizing human activities, especially using RF signals, is a challenge and open-source datasets would help in devising techniques and algorithms that can boost research in this field, which can ultimately lead to standardization.
A handful of works have made their datasets open-source which are intended for a number of applications such as WiFi CSI-based activity recognition \cite{ALAZRAI, Alsaify2020}, 
sign language recognition \cite{signfi}, fall detection \cite{FallDeFi}, device-to-device localization \cite{spotfi} or UWB-based gesture recognition \cite{UWB-gestures}, motion detection \cite{ klemenbregar5},  passive localization \cite{uwbmultistatic2020}, people counting \cite{uwbpeoplecount}, physical activity sensing using accelerometer and RSSI measurements from wearable sensors \cite{Byrne2018, GarciaGonzalez2020}, vision-based action recognition using Kinect \cite{kinectdataset}, while others have proposed radar-based simulators for generating synthetic datasets for day-to-day activities \cite{vishwakarma2021simhumalator}.
As compared to these works and to the best of the authors’ knowledge, this is the first work to propose a multimodal dataset comprising of RF and vision-based methods that is intended not only for sensing of day-to-day activities but also for passive (uncooperative) localization.
The contributions herein are:
\begin{itemize}
\item Multimodal data collection intended for human activity recognition and passive localization, i.e, the targets are oblivious to these processes (non-collaborative) and they only reflect or scatter the signals from the transmitter to receivers.
Most datasets consider only one particular modality such as either UWB, WiFi CSI, PWR or Kinect, independently. In this work, 
we consider multiple synchronized modalities. Experiments span across two environments which can be used for investigating sensing in complex or untrained environments.

\item Approximately 8 hours of measurements are annotated with high resolution location and activity labels, capturing the participant's movement and natural behaviour within the monitoring area, as would be the case in a real-world environment. The dataset is comprehensive in so far it contains over 1 Million annotated data points.


\item   The presented data can be exploited to advance human activity recognition technology in different ways, for example, using various pattern recognition and deep learning algorithms to accurately recognize human activities. For this purpose, the users can apply different signal processing pipelines to analyze the recorded WiFi CSI, PWR, UWB and Kinect data and extract salient features that can be used to recognize the human activities and/or concurrently track the target's position within an indoor environment.

 \item This is the first dataset that is collected with an explicit aim to accelerate the development of self-supervised learning techniques. Such techniques are extremely data hungry, requiring orders of magnitude larger datasets compared to more traditional supervised learning.

\end{itemize}

This open-source dataset is intended for both HAR and non-cooperative localization, which are areas of growing interest to research communities including but not limited to radar, wireless sensing, IoT and computer vision. 
To ensure that the dataset aligns to the FAIR (Findable, Accessible, Interoperable, Reusable) Data principles of Open Science, we have (i) made it publicly available for download via the figshare portal, (ii) provided an in-depth and clear description of the dataset for each modality, (iii) formatted our dataset using standard filetypes and encoding, and (iv) provided example scripts/codes that will allow the user to load and analyze the data from each modality.

\begin{figure}[ht]
	\begin{center}
		\includegraphics[width=1.1  \textwidth]{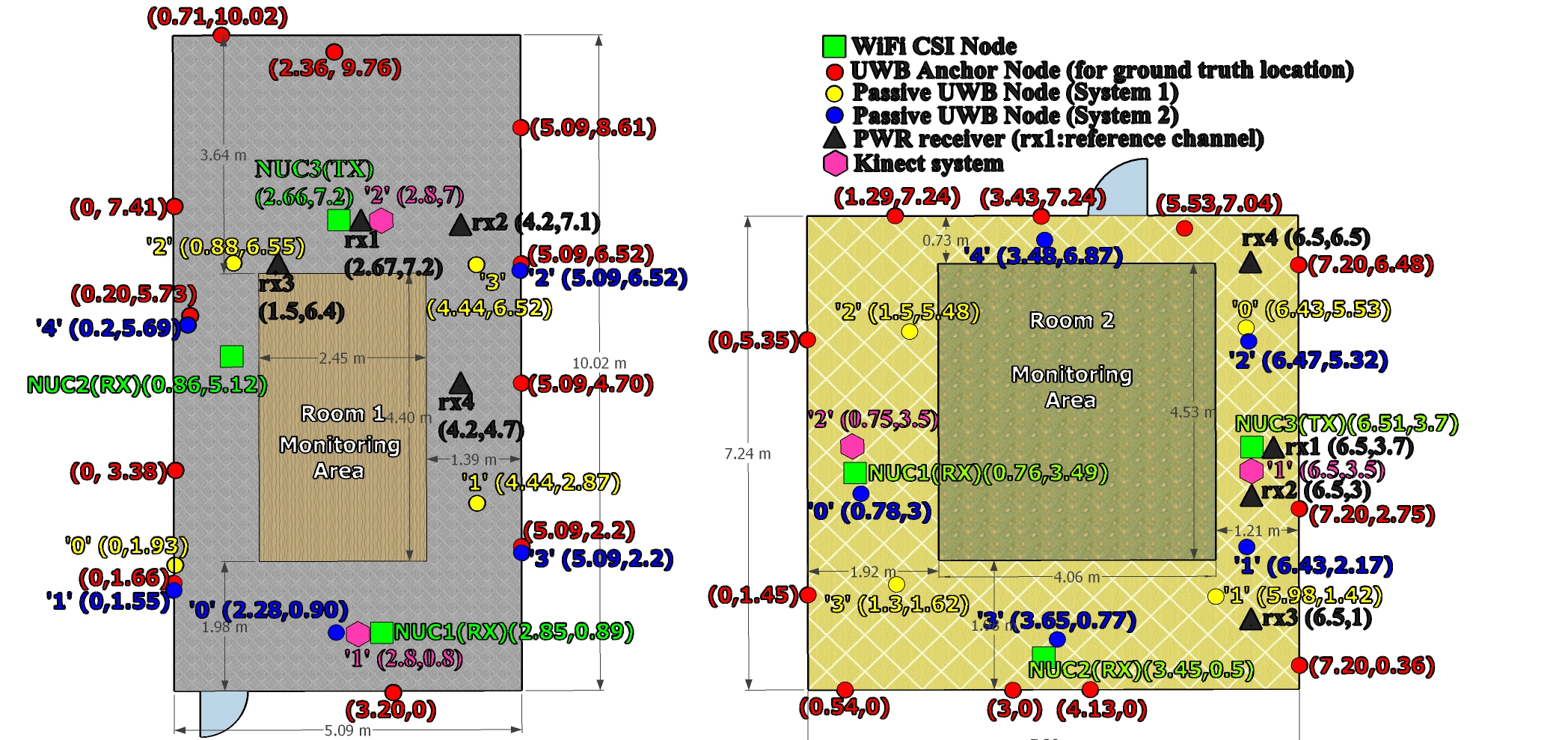}
\caption{Experiment rooms layouts.}
\label{room_layout}
	\end{center}
\end{figure}

\begin{table}[ht]
\centering
\begin{tabularx}{\columnwidth}{|X|X|}
\hline
Experiment No.  & Details  \\
\hline
\path{exp001, exp019, exp034,exp055} & Background (1) data (empty room).   \\
\hline
\path{exp002, exp006, exp010, exp014, exp020, exp024 } & Person walking (2).  \\
\hline
\path{exp003, exp007, exp011, exp015, exp021, exp025} & Person sitting (3) and standing from chair (4).        \\
\hline
\path{exp004, exp008, exp012, exp016, exp022, exp026} & Person lying down on floor (5) and standing up from floor (6).    \\
\hline
\path{exp005, exp009, exp013, exp017, exp023, exp027} &  Person rotating upper-half of his/her body (7).    \\
\hline
\path{exp018, exp029-exp033}  & Person performing the six activities (2-7) continuously and randomly (No predefined order).\\
\hline
\path{exp056-exp061}  & Person performing the six activities (2-7) in a predefined order, starting with activity "walking" and ending with activity "body rotating".\\
\hline
\path{exp028}  & Crowd counting. A maximum of six people walking continuously and randomly. Experiment starts with six people and then after every 5 min, one person steps out of the monitoring area.\\
\hline
\path{exp035-exp043}  & Device-free static localization. CSI transmitter (NUC3) and CSI receiver (NUC2) are placed side by side and the target stand still at a given position for each experiment number.\\
\hline
\path{exp044-exp048}  & Device-free dynamic localization. CSI transmitter (NUC3) and CSI receiver (NUC2) are placed side by side and the target moves along a short straight path for each experiment number.\\
\hline
\path{exp049-exp054}  & Device-to-device localization. No human target present. 
CSI transmitter (NUC3) and CSI receiver (NUC2) are placed at different angles with respect to each other (-30$^0$, 0$^0$, +30$^0$, -60$^0$, 0$^0$, +60$^0$) for each experiment number.\\
\hline
\end{tabularx}
\caption{\label{dataset_expno} Experiment description.}
\end{table}

\begin{table}[ht]
\centering
\begin{tabular}{|l|l|}
\hline
Activity  & Duration (min)  \\
\hline
Background & 18.3465  \\
\hline
Sit on chair& 35.2455  \\
\hline
Stand from chair & 34.8754  \\
\hline
Walk & 75.9323 \\
\hline
Lie down & 26.6891  \\
\hline
Stand from the floor & 26.4151  \\
\hline
Upper body rotate & 74.5893  \\
\hline
Steady state (no activity) & 124.8624  \\
\hline
Crowd counting & 27.4127  \\
\hline
Localization (CSI receiver NUC2 only) & 18.1803 \\
\hline
\end{tabular}
\caption{\label{activity_durations} Activity duration distribution.}
\end{table}



\begin{figure}[ht]
	\begin{center}
		\includegraphics[width=0.6 \textwidth]{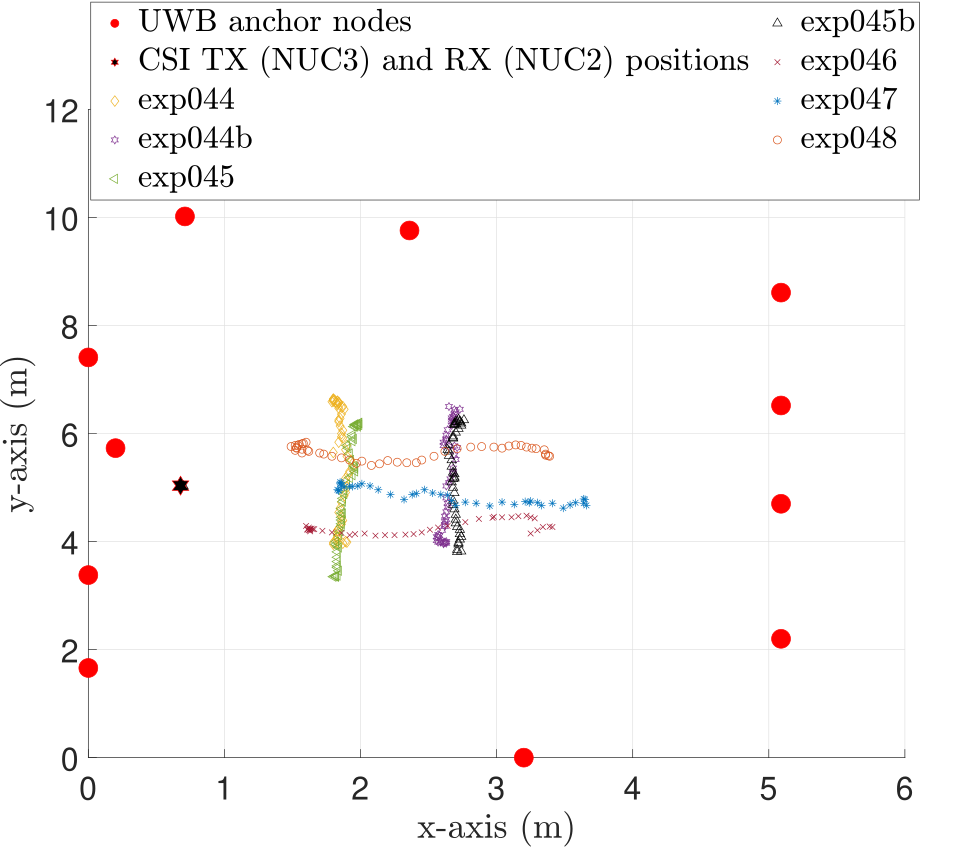}
\caption{Figure showing paths walked by target in dynamic localization experiments (\textbf{exp044-exp048}).}
\label{dyn_loc}
	\end{center}
\end{figure}

\begin{figure}[ht]
	\begin{center}
		\includegraphics[width=0.4  \textwidth]{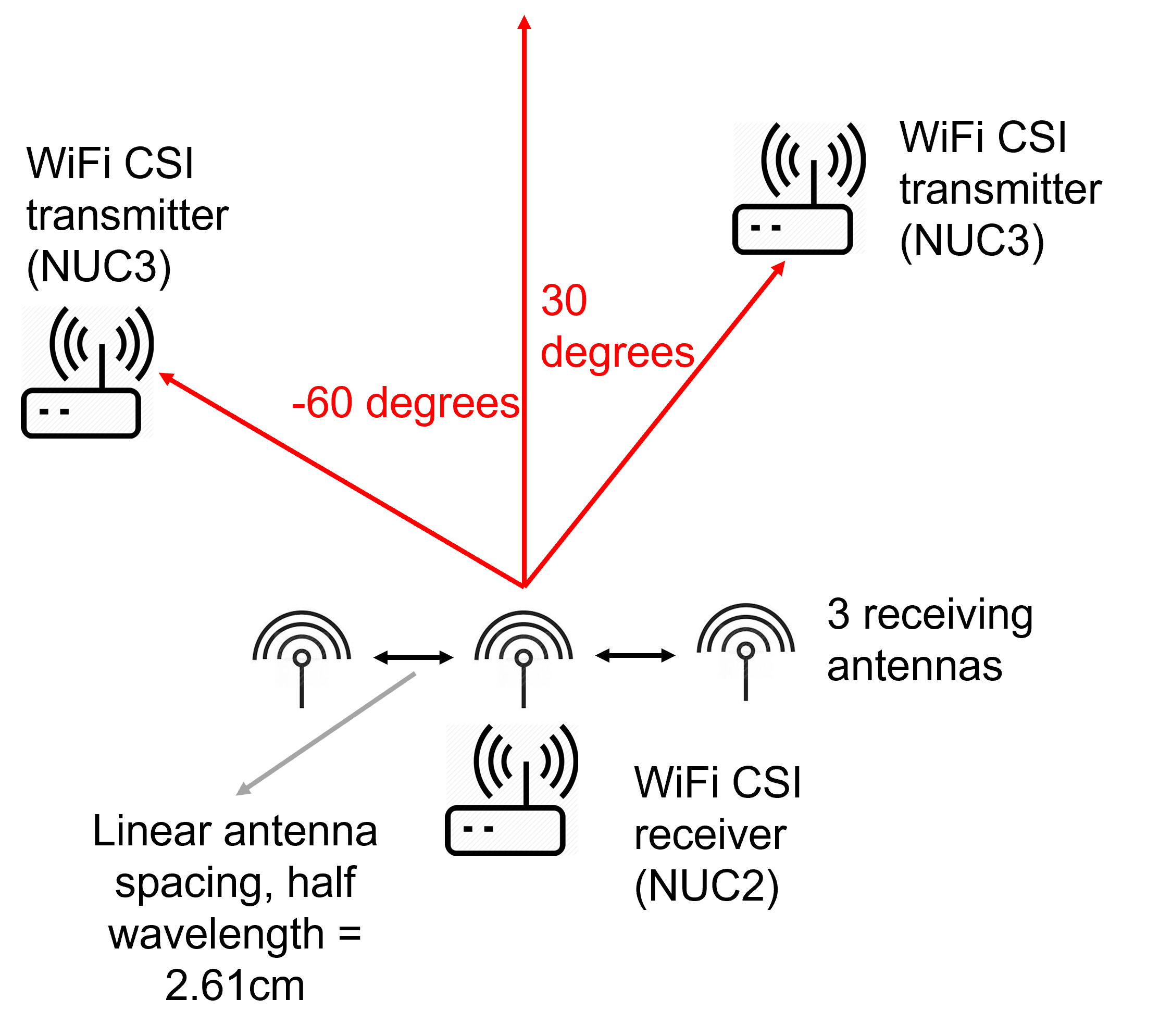}
\caption{Device-to-device localization experiment setup (\textbf{exp049-exp054}).}
\label{d2d_loc}
	\end{center}
\end{figure}

\section*{Methods}
Experiments were performed in a university environment in two furnished rooms, with desks, chairs, screens, and other office objects lying in the surroundings. The room layouts are depicted in Fig. \ref{room_layout} along with their physical dimensions. 
A maximum of six subjects of different age groups participated in the experiments which were intended for the sensing of day-to-day activities as well as non-collaborative localization. The description of the various experiments performed is provided in Table \ref{dataset_expno}. Approximately 8 hours of data were collected across multiple modalities including WiFi Channel State Information (CSI), Ultra-wideband (UWB), Passive WiFi Radar (PWR) and kinect sensor systems. The breakdown of the activities' durations is given in Table \ref{activity_durations}.
The monitoring devices were installed on the extremity (boundary) of the rooms such that enclosed spaces of dimensions 2.45m $\times$4.40m and 4.06m $\times$4.53m were used as monitoring areas for Room 1 and 2, respectively. 

Even though no personal data has been collected from the participants during the experiments, each participant was still fully informed about the purpose of the study and what was expected of them. Informed consent was obtained from each participant prior to the experiments. All studies that fall under the OPERA - Opportunistic Passive Radar for Non-Cooperative Contextual Sensing project were thoroughly reviewed and fully approved by the 
University of Bristol Faculty of Engineering Research Ethics Committee 
(application number: 96648). Risk assessment was also carried out and approved prior to the experiments.

\begin{figure}[ht]
	\begin{center}
		\includegraphics[width=0.4 \textwidth]{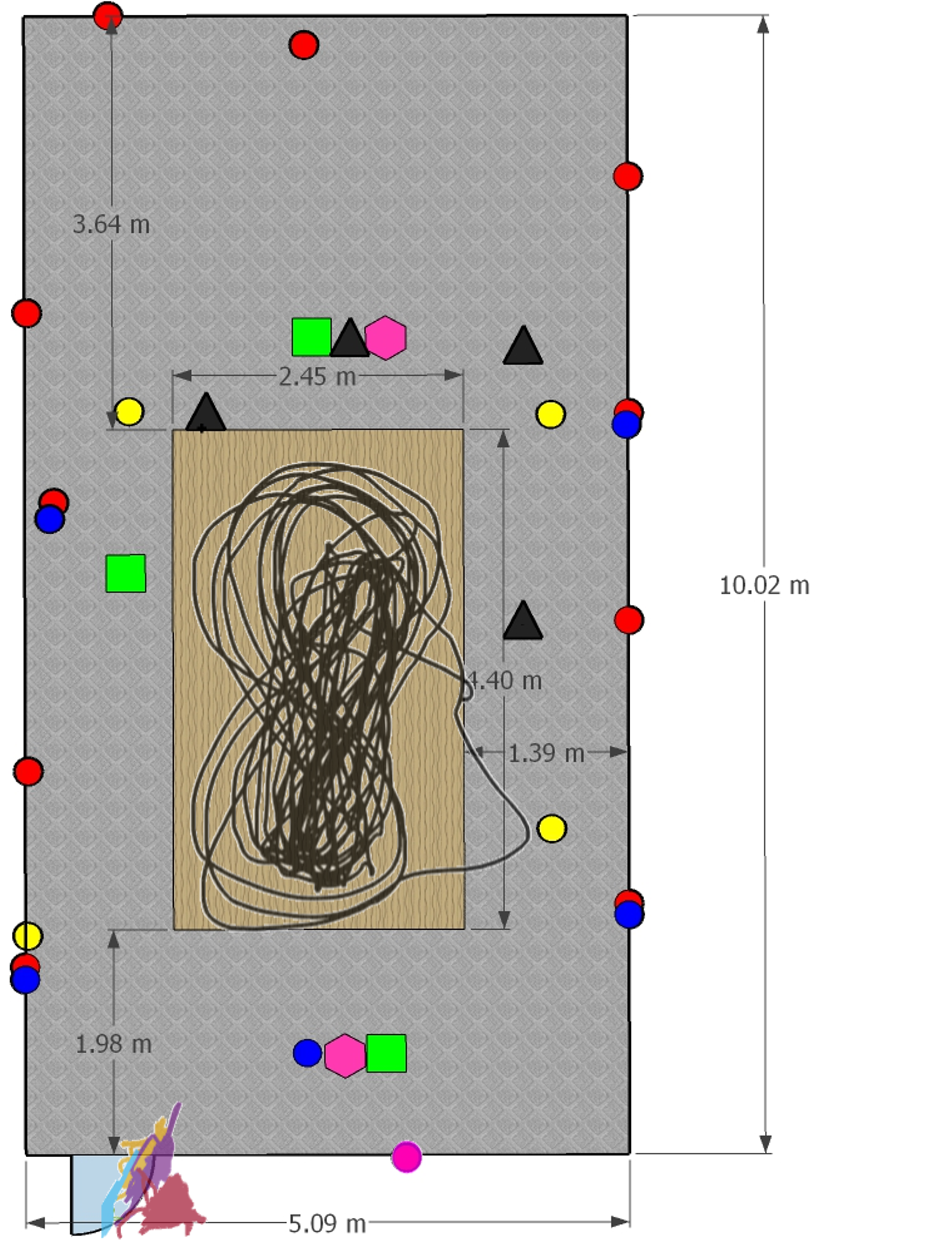} 
\caption{Crowd counting experiment (\textbf{exp028}) in Room 1. This picture shows the last person's walking path after the 5 previous participants have stepped out of the room sequentially.}
\label{cc_1person}
	\end{center}
\end{figure}

Referring to the experiment numbers in Table \ref{dataset_expno}, \path{exp001-exp054} were performed in Room 1 while \path{exp055-exp061} were carried out in Room 2.
\path{exp028} is the crowd counting experiment whereby six people walked randomly and continuously within the monitoring area of Room 1. Then, after approximately every 5 min, one person moved out of the room. 
Fig. \ref{cc_1person} shows the particular instant of \path{exp028} where 5 out of the 6 people already left the monitoring area and only the last person's ground truth walking trajectory is shown. For illustration purposes only, a moving average filter is applied to the raw ground truth positions to smooth the target's trajectory path.
The experiments \path{exp034-exp048} were device-free localization experiments involving a human target who was standing still at several positions or walking along a straight short path in a number of directions as shown in Fig. \ref{dyn_loc}. The target wore a tag to get his/her ground truth position. Note that only the WiFi CSI transmitter (NUC3) and receiver (NUC2) were used during these experiments for recording data and they were placed side by side. As for the device-to-device localization experiments (\path{exp049-exp054}), the CSI transmitter (NUC3) and receiver (NUC2) were placed at different angles with respect to each other (0$^0$, 30$^0$, -30$^0$, 60$^0$, -60$^0$ ), as shown in Fig. \ref{d2d_loc} (no human target). Note that one tag was placed on the CSI transmitter and another on the receiver to get their fixed ground locations within the environment.

\subsection*{WiFi CSI}
The WiFi CSI system consistsed of three Intel Core i5 vPro Next Unit of Computing (NUC) devices. Each device is fitted with an Intel5300 Network Interface Card (NIC). In order to extract the CSI from the NICs, the Linux 802.11n CSI tool \cite{csitool} need to be installed on the devices running an appropriate kernel version of the Linux operating system. Appropriate firmware and drivers need to be installed on the devices in order to expose the CSI. More information regarding the installation steps can be found in \cite{Linux80228}. The CSI provides information about the wireless channel characteristics such as multipath propagation, attenuation, phase shift, etc. It is regarded as a fine-grained information since it describes the amplitude and phase information of the signal across multiple overlapping but orthogonal subcarriers in the Orthogonal Frequency Division Multiplexing (OFDM) physical layer waveform. In a WiFi system based on OFDM, the CSI is used by the equalizer in the receiver to reverse the detrimental effects of the channel and recover the transmitted signal. The channel estimate (i.e., CSI) is obtained by transmitting a training sequence (pilot symbols) which is known by both the WiFi transmitter and receiver. This process is often referred to as channel sounding. CSI or Channel Frequency Response (CFR) are often used interchangeably and they represent the wireless channel in the frequency domain. Applying the Inverse Fast Fourier Transform (IFFT) to CFR gives rise to the Channel Impulse Response (CIR) in the time domain and this characterizes the amplitude and phase information over multiple propagation paths, as shown in Fig. \ref{uwb_cir}. 
The Intel 5300 NIC extracts CSI over 30 subcarriers, spread evenly among the 56 subcarriers of a 20 MHz WiFi channel or the 114 subcarriers in a 40 MHz channel \cite{csitool}. To record the CSI, the injector mode was used whereby one NUC was configured as the transmitter (injector) while the receivers were monitoring the channel into which packets were injected. This method requires that both the transmitter and receiver be equipped with the Intel5300 NIC. In the access point (AP) method, only the receiver needs to be equipped with the Intel5300 NIC and CSI data is logged at the receiver by pinging an access point. However, this method is not very stable since there might be a lot of dropped packets (e.g., due to interference) and also the packet rate is limited. The CSI data was stored on the receiver NUCs for offline processing. The raw data is in \path{.dat} format and they need to be parsed by appropriate Matlab utilities \cite{Linux80228} to obtain the data in a format that can be easily interpreted. Since the Intel5300 supports Multiple Input Multiple Output (MIMO) capability, the CSI data was logged as a 3D tensor of complex numbers for each received packet, with $n_t \times n_r \times N$ complex values, where $n_t$ is the number of transmit antennas, $n_r$ is the number of receive antennas and $N$ is the number of subcarriers. The parameters of the CSI system are summarized in Table \ref{csi_param}.
Referring to the green boxes in Fig. \ref{room_layout}, the WiFi CSI system  comprised of a single transmitter (NUC3) and two receivers (NUC1 and NUC2). NUC1 was installed facing the transmitter in a 
Line-of-Sight (LoS) geometry, while receiver NUC2 was placed in a bi-static configuration (90$^0$) with respect to the transmitter.

\begin{table}[ht]
\centering
\begin{tabular}{|l|l|l|}
\hline
System & WiFi CSI \\ \hline
WiFi Band & 5 GHz (channel 149)  \\ \hline
NIC & Intel5300    \\ \hline
Subcarriers, $N$  & 30 (out of 56)  \\ \hline
Antenna & omni-directional (6 dBi)  \\ \hline
Packet Rate & 1600 Hz \\ \hline
No. of transmit antennas, $n_t$ & 3  \\ \hline
No. of receive antennas,  $n_r$ & 3   \\ \hline
\end{tabular}
\caption{\label{csi_param} WiFi CSI system parameters.}
\end{table}

\subsection*{UWB}
Three UWB systems were used during the experiments. The first system (see red nodes in Fig. \ref{room_layout}) was used to obtain the ground truth position of the target while he/she wore one or more tags and moved within the monitoring area. The other two systems (yellow and blue nodes) were fixed nodes installed a multi-static configuration and which were exchanging Channel Impulse Response (CIR) data among themselves.
The 4 passive nodes of UWB system 1 (yellow nodes in Fig. \ref{room_layout}) were implemented using the Decawave's EVK1000 \cite{EVK1000} modules. The modules were programmed with a custom firmware so as to record CIR data on all modules. Node '0' was acting as an initiator whereby it exchanged Single-Sided Two-Way Ranging (SS-TWR) messages (poll, response and final) with each of the other 3 nodes. When a given node replies back, the frame is broadcast and heard by all other nodes operating on the same channel. In this way, each node can read the received frames in their accumulator and extract the CIR data.
Therefore, CIR data is available in a bidirectional mode between all pairs of nodes. This means that all nodes act as transmitters and receivers, giving rise to a maximum of 12 communication channels. The 4 nodes were connected to laptops in order to record the CIR data via a serial terminal. 

The 5 passive nodes of UWB system 2 (blue nodes in Fig. \ref{room_layout}) were implemented using the Decawave's MDEK1001 \cite{MDEK100157} modules. The units were flashed with custom firmware so as to record CIR data on all nodes. Node '0' was acting as an initiator and transmitted a packet at a set period. The packet essentially includes a time schedule for transmission for the other 4 nodes. In this way, each node knows who needs to transmit next and when with minimal delay. Thus the transmission was performed in a round-robin fashion to avoid collision. Nodes with IDs 1-4 were connected to laptops to record the CIR data via a serial terminal. The average packet rate for UWB system 1 (yellow nodes) was around 400Hz 
while for UWB system 2 (blue nodes), the average packet rate was around 
195Hz,
considering combined communication links. The other parameters for UWB system 1 and 2 are summarized in Table \ref{uwb_param}.

\begin{table}[ht]
\centering
\begin{tabular}{|l|l|l|}
\hline
Parameter               & UWB system 1 &  UWB system 2 \\ \hline
Channel          & 4             &   3             \\ \hline
Carrier Frequency&  3993.6 MHz             &     4492.8 MHz          \\ \hline
Bandwith         &  1331.2$^*$ MHz             &   499.2 MHz           \\ \hline
Pulse Repetition Frequency  & 64 MHz         & 64 MHz \\ \hline
Data Rate         &  6.8 Mbps        &   6.8 Mbps\\ \hline
Preamble Length	& 128 Symbols   & 128 Symbols\\ \hline
Preamble Accumulation Size   &   8       & 8 \\ \hline
Preamble Code              &    17      & 9 \\ \hline
\end{tabular}
\caption{\label{uwb_param} UWB systems' parameters ($^*$maximum receiver bandwidth is approximately 900 MHz).}
\end{table}

The Decawaves's UWB chipset stores the CIR in an accumulator and
each tap of the accumulator represents a sampling interval of $\Delta \tau_{\text{s}}$$\approx$1.0016 ns (i.e., half a period of the 499.2 MHz fundamental frequency)\cite{DW1000ma}. 
The accumulator spans over one symbol time. This represents 992 and 1016 samples for the nominal Pulse Repetition Frequency (PRF) of 16 MHz and 64 MHz, respectively.
Each measured CIR sample is a complex number which can be broken down into its real and imaginary components.
Only 35 and 50 CIR samples out of 1016 are considered in the experiments for UWB systems 1 and 2, respectively. These correspond to a sensing range of 10.5m and 15m for UWB systems 1 and 2, respectively.
Each CIR measurement was read from the accumulator memory starting 3 samples (i.e., 3 ns) before the detected first path index (reported by the \path{FP_INDEX} field in register \path{0x15} of DW1000 chipset) by the LDE algorithm. 
As the CIR magnitudes are dependent on the number of preamble symbols used for CIR accumulation, for each CIR measurement in the UWB datasets, the magnitude values are normalized using the 
Preamble Accumulation Count (PAC)
value \cite{safetycir2016} (see \path{rx_pream_count} column in UWB datasets), as reported by the \path{RXPACC} register in the DW1000 chip.


\begin{table}[ht]
\centering
\begin{tabular}{|l|l|l|}
\hline
System & WiFi PWR \\ \hline
WiFi Band & 5 GHz (channel 149), the CSI transmitter \\ \hline
RF Frontend & USRP 2945    \\ \hline
Antenna & omni-directional (6 dBi)  \\ \hline
Packet Rate & 1600 Hz \\ \hline
No. of surveillance channel, $n_s$ & 3  \\ \hline
No. of reference channel, $n_r$ & 1   \\ \hline
Number of Doppler bins, $n_b$ & 200 \\ \hline
\end{tabular}
\caption{\label{PWR_param} PWR system parameters.}
\end{table}

\subsection*{PWR}
For the Passive WiFi radar (PWR) system, a USRP-2945 \cite{USRP2945} was used as the receiver which is equipped with four synchronized channels. 
The USRP-2945 features a two-stage superheterodyne architecture with four independent receiving channels and shares local oscillators for phase-coherent operation. 
Each receiving channel was equipped with a 6-dB directional antenna. 
The collected raw data are then routed to a computing unit through a PCIe port, which is a desktop computer in this work. 
A PWR system consists of a minimum of two synchronized channels; a surveillance channel which records reflected WiFi signals from the monitoring area and a reference channel which records the direct signal emitted from the transmitter. 
As mentioned previously, four channels are used in the USRP-2945, where one channel was used as the reference channel (denoted as "rx1" in Fig. \ref{room_layout}) while the other three channels were used as surveillance channels (denoted as "rx2", "rx3" and "rx4" in Fig. \ref{room_layout}). 
Since PWR does not transmit a signal (it only monitors received signals), it 
can use
any third-party signal source as the illuminator but however, a reference signal is needed. 
In this work, we used the CSI transmitter (NUC3) as the PWR source for convenience, 
allowing
a direct comparison between the two systems' performance. 

PWR correlates the signal from the surveillance and reference channels to estimate two parameters: relative range and Doppler frequency shift. 
Additionally, a CLEAN \cite{li2020passive} algorithm has been used to remove the direct signal interference. More details on this signal processing can be found at \cite{li2020passive}. 
However, due to the limitation of the WiFi signal bandwidth (40MHz in this work), the range resolution is limited to 3.75 meters which is too coarse for indoor applications. 
Therefore, only the Doppler frequency shifts are recorded in the form of Doppler spectrograms. 
The output from the PWR system is specified as $n_s \times n_b \times N_t $ real values, where $n_s$ is the number of surveillance channels, $n_b$ is the number of Doppler bins and $N_t$ is the time. 
Other details about the PWR's parameters are given in Table \ref{PWR_param}. 

\subsection*{Kinect}
We used two of Microsoft's Kinect v2 sensors to gather motion capture data from different human activities \cite{Kinect}. Kinect v2 incorporates an Infrared depth sensor, a RGB camera, and a four-element microphone array that provides functionalities such as three-dimensional skeletal tracking, facial recognition, and voice recognition. Although the device was originally developed to play games, numerous researchers have used it for applications beyond its initial intended purpose. Due to the low cost and wide availability, it has now been used extensively in research areas, such as video surveillance systems where multiple Kinect devices are synchronized to track groups of people even in complete darkness \cite{chai2012general}, improve live three-dimensional videoconferencing \cite{kramer2012hacking} and in medicines to measure a range of conditions such as autism, attention-deficit disorder and obsessive-compulsive disorder in children \cite{nyberg2014development}. Note that in skeleton tracking, Kinect might suffer from occlusion when some parts of the human body are occluded with others and therefore cannot be tracked accurately. 

Therefore, in this work, we used two Kinects to track three-dimensional time-varying skeletal information of the human body, including 25 joints such as head center location, knee joints, elbow joints, and shoulder joints from two different directions. 
The real advantage of using motion capture technology is capturing more accurate, more realistic, and complex human motions. This three-dimensional joint information can further be used for simulating the corresponding radar scatterings mimicking a typical PWR sensing system. In one of our previous works, we presented an open-source motion capture data-driven simulation tool, \textit{SimHumalator}, that can generate large volumes of human micro-Doppler radar data at multiple IEEE WiFi standards (IEEE 802.11g, ax, and ad) \cite{vishwakarma2021simhumalator}. Radar scatterings were simulated by integrating the animation data of humans with IEEE 802.11 complaint WiFi transmissions to capture features that incorporate the diversity of human motion characteristics and the sensor parameters. More importantly, we have demonstrated that the human micro-Doppler data generated using the simulator can be used to augment limited experimental data \cite{tang2021augmenting,vishwakarma2021gan}. Interested researchers can download the simulator from \url{https://uwsl.co.uk/}.

The output from the Kinect system is specified as $N_t \times N_b \times N_d $ real values, where $N_t$ is the number of time frames, $N_b$ is the number of tracked joints on the human body, and $N_d$ is the three-dimensional position ($x,y,z$) information. 
 \subsection*{Ground Truthing}
 \begin{itemize}
\item The Decawave's (now Qorvo) MDEK1001 development kit \cite{MDEK100157} was used for obtaining the ground truth position of the targets. 11 units were configured as anchors and mounted on walls in the experiment rooms (see red nodes in Fig. \ref{room_layout}). Their xy coordinates were manually measured using a laser measuring device, which were then entered in the DRTLS Android app. A maximum of 6 tags were configured for \path{exp028}, while for the activity recognition experiments, the person wore two tags, one on each arm. Two UWB units were also configured as listeners so as to record the xy coordinates of the tags on a serial terminal on two laptops (for redundancy). 
 
\item As for the labelling of activities, a program was developed in Matlab with automated voice output to instruct the person when to perform the various activities such as sitting, standing, etc. At the same time, the programmable script recorded the timestamps at which the activity was instructed to be performed. The person just had to listen to the voice command and perform the activity accordingly. As a backup solution, another activity labelling application was developed in Matlab where one can insert the labels for the required activities. Then, an observer constantly looked at the person doing the activities and clicked on the appropriate button in the app to record the start and stop times of the activity. All labels were stored in text files along with their timestamps. Note that all modalities were synchronized to the same local Network Time Protocol (NTP) server, resulting in synchronisation accuracy across all modalities of < 20ms. 
\end{itemize}

\section*{Data Records}

Four modalities have been used during the experiments, namely, WiFi CSI, UWB, PWR and Kinect sensor. With respect to WiFi CSI and UWB, there were two systems in each case. 
The experimental datasets can be accessed and downloaded from the figshare repository at \url{https://figshare.com/s/c774748e127dcdecc667}. The datasets have been compressed (zipped) into separate folders for each modality, allowing the user to only download the data of interest. The zipped folders' names and the number of files in each folder, along with their file formats, are specified in      
Table \ref{dataset_direc}. The directories \path{wificsi1} and \path{wificsi2} refer to the data collected by the WiFi CSI receivers, denoted by "NUC1" and "NUC2" in Fig. \ref{room_layout}, respectively. \path{uwb1} and \path{uwb2} refer to the data collected by the two passive UWB systems, represented by the yellow and blue nodes in Fig. \ref{room_layout}, respectively. 
The directory \path{pwr} contains the PWR spectrogram data recorded from the three surveillance channels ("rx2", rx3" and "rx4" represented as black triangles in Fig. \ref{room_layout}) for each experiment (excluding \path{exp001,exp019}, \path{exp034-exp054}). Finally, the directory \path{kinect} contains the Kinect sensor data files for each experiment (excluding \path{exp001,exp019}, \path{exp034-exp054}, \path{exp055}).


\subsection*{Experiment Directory}
Each file in the directories in Table \ref{dataset_direc} corresponds to a given experiment number (filename contains strings \path{exp001,exp002}, etc.,), the details of which are provided in Table \ref{dataset_expno}.

\begin{table}[ht]
\centering
\begin{tabular}{|l|l|l|}
\hline
Directory Name  & No. of files & File format  \\
\hline
\path{wificsi1} & 40 & .mat  \\
\hline
\path{wificsi2} & 63 & .mat  \\
\hline
\path{uwb1} & 40    & .csv      \\
\hline
\path{uwb2} & 40   & .csv     \\
\hline
\path{pwr} &  38 &  .mat     \\
\hline
\path{kinect}  &  36    &  .mat\\
\hline
\end{tabular}
\caption{\label{dataset_direc} Dataset directory details.}
\end{table}

\subsubsection*{WiFi CSI dataset description}
This section describes the structure of the data files residing in the \path{wificsi1} (NUC1) and \path{wificsi2} (NUC2) directories. The files are in \path{.mat} format and each row in the file corresponds to a received CSI packet. The columns in the dataset have the following headers:

\begin{itemize}
\item  \path{timestamp}: UTC+01 00 timestamp in milliseconds when the CSI packet was received by the NUC devices.

\item  \path{activity}: current activity being performed. The activity is specified as a string of characters with no spacing e.g., "background", "walk", "sit", "stand", "liedown", "standfromlie", "bodyrotate". These correspond to the activity numbers 1, 2, 3, 4, 5, 6 and 7 in the "Details" column in Table \ref{dataset_expno}, respectively. 
The activity label "noactivity" refers to the case where the person was not performing any activity, that is, his/her body was at rest, for example between activities such as "sitting" and "standing" or "lying down on floor" and "standing up from the floor". 
For \path{exp035-exp054}, the activity is specified as "Loc1", "Loc2", $\cdots$, "Loc9" (device-free static target localization), "path1", "path2", $\cdots$, "path5" (device-free dynamic target localization), "loc1", "loc2", $\cdots$, "loc6" (device-to-device localization). For these localization experiments, each file also contains a column with header \path{notes} which gives more details on the position of the transmitter (tx) and receiver (rx) and the target (if applicable).

\item \path{exp_no}: experiment number which is specified as "exp\_001", "exp\_002", etc. See Table \ref{dataset_expno} for more details.

\item \path{person_id}: person ID specified as "One", "Two", "Three", etc.

\item \path{room_no}: room ID specified as "1" (left room in Fig. \ref{room_layout}) or "2" (right room in Fig. \ref{room_layout}). 

\item \path{tag4422_x}, \path{tag4422_y}, \path{tag89b3_x}, \path{tag89b3_y}: refer to the ground truth position of the person in the monitoring area in terms of 2D x- and y- coordinates. Note that for all experiments, except \path{exp001, exp019, exp034, exp055, exp028 } and \path{exp035-exp054} (NUC2 only), the person was wearing two UWB tags on either arms, bearing IDs \path{4422} and \path{89b3}. The information regarding which tag is worn on which arm is given in the columns with headers \path{left_arm_tag_id} and \path{right_arm_tag_id}. For the crowd counting experiment (\path{exp028}), there were a maximum of 6 people and hence 6 UWB tags were used to obtain the ground truth position of each person. Each person wore the tag on his/her left arm. In the WiFi CSI files for \path{exp028}, the x- and y- coordinates of the person is given in the columns \path{tag4422_x}, \path{tag4422_y}, \path{tag89b3_x}, \path{tag89b3_y}, \path{tag122c_x}, \path{tag122c_y}, \path{tag4956_x}, \path{tag4956_y}, \path{tag1e85_x}, \path{tag1e85_y}, and \path{tag9118_x}, \path{9118_y}.
The person bearing UWB tag id \path{4956} was the first to step out of the monitoring area, followed by \path{9118}, \path{1E85}, \path{4422}, \path{89B3}, and finally \path{122C}.

\item \path{anchor_node_xy_positions}: x- and y- coordinates of the eleven UWB anchor nodes distributed across the room (see red nodes in Fig. \ref{room_layout}) for obtaining the ground truth position of the tag/s.

\item \path{tx1rx1_sub1, tx1rx1_sub2}, $\cdots$, \path{tx3rx3_sub30} (270 columns): The first corresponds to the raw complex CSI values for transmit antenna 1 (\path{tx1}), receive antenna 1 (\path{rx1}) and subcarrier 1 (\path{sub1}), the second corresponds to transmit antenna 1 (\path{tx1}), receive antenna 1 (\path{rx1}) and subcarrier 2 (\path{sub2}), and so on. The WiFi CSI systems used a 3$\times$3 MIMO configuration and since the Intel5300 NIC extracts CSI data over 30 subcarriers, the total number of complex CSI values per packet is 3$\times$3$\times$30=270.

\item \path{tx_x_coord}, \path{tx_y_coord}, \path{target_x_coord}, \path{target_y_coord} (for \path{exp035-exp054} only): For \path{exp035} -\path{exp048}, \path{tx_x_coord} and \path{tx_y_coord} respectively correspond to the x- and y- coordinates of both the CSI transmitter (NUC3) and CSI receiver (NUC2) since they were placed side by side while the target was standing still at several positions or walking along a short path. The human target was holding a tag and its ground truth x- and y- coordinates are given by \path{target_x_coord} and \path{target_y_coord}, respectively.
As for \path{exp049} -\path{exp054}, \path{tx_x_coord} and \path{tx_y_coord} refer to the x- and y- coordinates of the CSI transmitter (NUC3), respectively, while \path{target_x_coord} and \path{target_y_coord} refer to the x- and y- coordinates of the CSI receiver (NUC2), respectively. No human target was present in this case.

\end{itemize}

\subsubsection*{UWB dataset description}
This section describes the structure of the data files residing in the \path{uwb1} (UWB system 1- yellow nodes) and \path{uwb2} (UWB system 2- blue nodes) directories. The files are in \path{.csv} format and each row in the file corresponds to a received UWB packet.
The UWB dataset files have the following fields similar to the WiFi CSI datasets: \path{timestamp}, \path{activity}, \path{exp_no}, \path{person_id}, \path{room_no}, \path{tag4422_x}, \path{tag4422_y}, \path{tag89b3_x}, \path{tag89b3_y}, \path{tag122c_x}, \path{tag122c_y}, \path{tag4956_x}, \path{tag4956_y}, \path{tag1e85_x}, \path{tag1e85_y}, \path{tag9118_x}, \path{9118_y}, \path{left_arm_tag_id},  \path{right_arm_tag_id}, and \path{anchor_node_xy_positions}. The additional column headers or those that are different from the WiFi CSI dataset headers are described below:

\begin{itemize}
\item  \path{fp_pow_dbm}:  estimate of the first path power level (in dBm) of the UWB signal between a pair of nodes. The formula for computing this value is given in \cite{DW1000ma}.

\item  \path{rx_pow_dbm}:  estimate of the receive power level (in dBm) of the UWB signal between a pair of nodes. The formula for computing this value is given in \cite{DW1000ma}. 

According to the manufacturer, the above two estimated parameters can be used to infer whether the received signal is Line-of-Sight (LoS) or Non-Line-of-Sight (NLOS). It is stated that, as a rule of thumb, if the difference of the two parameters, i.e., \path{rx_pow_dbm} - \path{fp_pow_dbm} is less than 6dB, the signal is most likely to be LoS, whilst if the difference is greater than
10dB, the signal is likely to be NLoS \cite{DW1000ma}.

\item  \path{tx_id}: index of the transmitting node. For UWB system 1 (yellow nodes), the transmitting node IDs are 0, 1, 2 or 3. For UWB system 2 (blue nodes), the transmitting node IDs are 0, 1, 2, 3 or 4 (see Fig. \ref{room_layout}).

\item  \path{rx_id}: index of the receiving node. For UWB system 1 (yellow nodes), the receiving node IDs are 0, 1, 2 or 3. For UWB system 2 (blue nodes), the receiving node IDs are 1, 2, 3 or 4.

\item  \path{tx_x_coord,tx_y_coord}: x- and y- coordinates of the transmitting node, respectively.

\item  \path{rx_x_coord,rx_y_coord}: x- and y- coordinates of the receiving node, respectively.

\item  \path{tx_rx_dist_meters}: separation distance between the pair of transmitting and receiving nodes in meters.

\item  \path{fp_index}: accumulator first path index as reported by the Leading Edge Detection (LDE) algorithm of the DW1000 UWB chipset in register \path{0x15} (in \path{FP_INDEX} field). It is a sub-nanosecond quantity, consisting of an integer part and a fractional part.

\item  \path{fp_amp1}: first path amplitude (point 3) value reported in the \path{FP_AMPL1} field of register 0x15 of the DW1000 UWB chipset. 

\item  \path{fp_amp2}: first path Amplitude (point 2) value reported in the \path{FP_AMPL2} field of register \path{0x12} of the DW1000 UWB chipset. 

\item  \path{fp_amp3}:  first path amplitude (point 1) value reported in the \path{FP_AMPL3} field of register \path{0x12} of the DW1000 UWB chipset. 

Basically, \path{fp_amp1}, \path{fp_amp2} and \path{fp_amp3} are the magnitudes of the accumulator tap at the indices 3, 2 and 1, respectively, beyond the integer part of \path{FP_INDEX} reported in register \path{0x15} of the DW1000 UWB chipset \cite{DW1000ma}. That is, \path{fp_amp1} = amplitude at ceiling(\path{FP_INDEX}) + 3, \path{fp_amp2} = amplitude at ceiling(\path{FP_INDEX}) + 2 and \path{fp_amp3} = amplitude at ceiling(\path{FP_INDEX}) + 1.

\item  \path{max_growth_cir}: Channel Impulse Response (CIR) power value reported in the \path{CIR_PWR} field of register \path{0x12} of the DW1000 UWB chipset. This value is the sum of the squares of the magnitudes of the accumulator from the estimated highest power portion of the channel, which is related to the receive signal power \cite{DW1000ma}.

\item  \path{rx_pream_count}: Preamble Accumulation Count (PAC) value reported in the \path{RXPACC} field of register \path{0x10} of the DW1000 UWB chipset. \path{RXPACC} reports the number of accumulated preamble symbols. The DW1000 chip estimates the CIR by correlating a known preamble sequence with the received signal and accumulating the
result over a time period. The number of preambles used for the CIR estimation is dependent on the quality of the received signal.

\item  \path{max_noise}: LDE max value of noise.

\item  \path{std_noise}: standard deviation of noise.

\item  \path{cir1, cir2}, $\cdots$: These correspond to the 35 and 50 raw complex CIR samples for UWB systems 1 and 2, respectively.
\end{itemize}


\subsubsection*{PWR dataset description}
This section describes the structure of the data files residing in the \path{pwr} directory.
The files are in \path{.mat} format and each row in the files corresponds to a PWR measurement from each of the three receivers (surveillance channels) at a given point in time.

\begin{itemize}
\item  \path{exp_no}: experiment number which is specified as "exp\_002", "exp\_003", etc. See Table \ref{dataset_expno} for more details. Note that the PWR system does not need background scan. Hence, background data for "exp\_001" and "exp\_019" are omitted for the PWR system.   

\item  \path{timestamp}: UTC+01 00 timestamp in milliseconds when the Doppler spectrograms were recorded. 

\item  \path{activity}: ground truth activity labels. The activity is specified as a string of characters with no spacing e.g., "walk", "sit", "stand", "liedown", "standfromlie", "bodyrotate". These correspond to the activity numbers 1, 2, 3, 4, 5, 6 and 7 in the "Details" column in Table \ref{dataset_expno}, respectively. 

\item  \path{person_id}: person ID specified as "One", "Two", "Three", etc.

\item  \path{room_no}: room ID specified as "1" (left room in Fig. \ref{room_layout}) or "2" (right room in Fig. \ref{room_layout}).

\item  \path{PWR_ch1}: Doppler spectrogram measured from surveillance channel "rx2", as demonstrated in Fig. \ref{signals_csi_uwb}(d). 

\item  \path{PWR_ch2}: Doppler spectrogram measured from surveillance channel "rx3".

\item  \path{PWR_ch3}: Doppler spectrogram measured from surveillance channel "rx4".

\end{itemize}

\subsubsection*{Kinect dataset description}
This section describes the structure of the data files residing in the \path{kinect} directory.
The files are in \path{.mat} format and each row in the files corresponds to three-dimensional skeleton information captured from each of the two Kinects at a given point in time.

\begin{itemize}
\item  \path{exp_no}: experiment number which is specified as "exp\_002", "exp\_003", etc. See Table \ref{dataset_expno} for more details. Note that the Kinect system does not need background scan. Hence, background data for "exp\_001" and "exp\_019" are omitted for the Kinect data.   

\item  \path{timestamp}: UTC+01 00 timestamp in milliseconds when the Kinect skeleton data were recorded. 

\item  \path{activity}: ground truth activity labels. The activity is specified as a string of characters with no spacing e.g., "walk", "sit", "stand", "liedown", "standfromlie", "bodyrotate". These correspond to the activity numbers 1, 2, 3, 4, 5, 6 and 7 in the "Details" column in Table \ref{dataset_expno}, respectively. 

\item  \path{person_id}: person ID specified as "One", "Two", "Three", etc.

\item  \path{room_no}: room ID specified as "1" (left room in Fig. \ref{room_layout}) or "2" (right room in Fig. \ref{room_layout}).

\item  \path{Kinect1}: Velocity-time profile measured from Kinect skeleton data over a period of time is demonstrated in Fig. \ref{signals_csi_uwb}(c). It perfectly captures human motion characteristics and is qualitatively similar to the envelope of the human-micro-Doppler signatures presented in \ref{signals_csi_uwb}(d).
\end{itemize}

\section*{Technical Validation}

\subsection*{WiFi CSI}
Fig. \ref{signals_csi_uwb}(a) shows 
a 196s portion
of the received WiFi CSI signals on NUC1 and NUC2 for \path{exp018}. CSI values for transmit antenna 1, receive antenna 1 and subcarrier 10 has been considered here. The injection packet rate of WiFi CSI was set a 1600 Hz. 
For illustration purposes only, the CSI data has been filtered using
a 1D wavelet denoising technique 
and the corresponding results are shown in Fig. \ref{signals_csi_uwb}(a). It can be observed that the CSI  measurements in the  time  domain  capture  variations in  the  wireless  signal  due  to  the  latter's  interaction with surrounding  objects and human  bodies. Therefore, machine or deep learning algorithms can be used to train the observed patterns and automatically extract features from raw signals to predict human activities. The CSI data can be processed and interpreted in different ways (feature extraction), for example spectrograms which are generated by applying Short Time Fourier Transform (STFT) to the CSI amplitude data \cite{translationresilientcsi,taxonomycsi, bocus2021uwb}.
Note that the two receiving NUCs (NUC1 and NUC2) were arranged differently with respect to the transmitter (NUC3). As shown in Fig. \ref{room_layout}, in both rooms NUC1 was facing the transmitter in a 180$^0$ configuration while NUC2 was in a bistatic geometry (90$^0$) with respect to the transmitter. By having multiple views of the same activity being performed, it is envisaged that the prediction accuracy can be improved using algorithms such as contrastive learning \cite{lau2021selfsupervised}.

\subsection*{UWB}
Fig. \ref{signals_csi_uwb}(b) shows the UWB signals between nodes '0' and node '3' for UWB system 1 and nodes '1' and '2' for UWB system 2, considering the same experiment number and time window. The raw CIR data has been converted to Channel Frequency Response (CFR) using the Fast Fourier Transform (FFT) and the signals are plotted for the $10^{th}$ CFR sample for each system. Note that the terms CFR and CSI can be used interchangeably. Although the sampling rate of the UWB systems is much lower (typically <100 Hz considering bidirectional data which are reciprocal) than the WiFi CSI system, the activities cause variations in the UWB signals and these variations can be fed to machine or deep learning algorithms for activity prediction. The raw CIR can also be used for human activity recognition (HAR) and yield high prediction accuracy, as demonstrated in \cite{bocus2021uwb}.

Considering the crowd counting experiment, Fig. \ref{fp_pow_figa} and \ref{fp_pow_figb}  show the first path power level (in dBm) for the two UWB systems between a given pair of nodes in each case. The first path power level has been computed using the formula given in the DW1000 manual \cite{DW1000ma}. As can be observed, the first path power level increases gradually as each person was moving out of the monitoring area. This is an expected behaviour since the LoS signal becomes less and less obstructed. By using the \path{fp_pow_dbm} parameter together with other parameters such as overall received UWB power level (\path{rx_pow_dbm}), UWB CIR data and WiFi CSI data, the number of people in a given environment can be inferred through the use of artificial intelligence algorithms.  

While UWB modules such as Decawave's EVK1000 and DWM1000 are used for active localization providing the 2D or 3D coordinates of a target carrying a tag, in this experiment, we deployed fixed UWB nodes programmed with custom software to record CIR data in a multi-static configuration. The idea is to extend the functionality of pulse-based UWB systems from active localization to the ability to sense their environment using the CIR data \cite{safetycir2016,uwbmultistatic2020}.
Fig. \ref{uwb_cira} and \ref{uwb_cirb} show 1000 aligned and accumulated CIR measurements recorded between a given pair of UWB modules for each system in a static environment (\path{exp001}). As can be observed from Fig. \ref{uwb_cira} and \ref{uwb_cirb}, when the room is empty, the accumulated CIR measurements are stable. However, when a person is performing activities as in \path{exp003}, some variations occur in the accumulated CIR, as can be observed in the region starting around $\tau-\tau_{\text{FP}}$ $\approx$ 8ns and 10ns in Fig. \ref{uwb_circ} and \ref{uwb_cird}, respectively.
The earliest time at which changes are observed in the CIR is the bi-static delay. Since the transceivers are fixed in the multi-static network and their positions are known, the distance travelled by the direct (first path) signal between pairs of devices can be computed along with its delay $\tau_{\text{FP}}$. The black vertical lines in Fig. \ref{uwb_circ} and \ref{uwb_cird} represent the ground truth bi-static delay which is computed from the tags' coordinates. Note that the target was wearing two tags, one on each arm. Also, \path{exp003} refers to the sitting on chair and standing up from chair activities which were performed at a given location within the monitoring area. Therefore, the reported locations for each tag were averaged over the 1000 accumulated CIR measurements to obtain a single 2D position per tag. Now, the separation between the two tag' coordinates are of a certain distance apart, which is around 60cm, corresponding to the approximate diameter of the human (arm to arm distance). Therefore, the midpoint xy coordinates are taken as the ground truth position of the target. The ground truth bi-static delay can then be computed by finding the transmitter-target-receiver path length and subtracting the direct signal path length from it, given that the fixed UWB transmitter and receiver positions are known.
Assuming that the signal emitted from the transmitter reflects off the target and reaches the receiver without additional scattering, then the bi-static range defines an ellipse on which the target is located \cite{uwbmultistatic2020}. This ellipse has the position of the transmitter and receiver as foci points and the major axis length is equal to the bi-static range. In an ideal scenario, the common intersection point of ellipses from multiple transmitter and receiver pairs indicates the location of the target.


It should be noted that in the multi-static UWB network, each transceiver device runs its own independent RF clock 
and therefore CIR measurements between pairs of devices may be sampled at different times \cite{uwbmultistatic2020}. 
The DW1000 chip organizes the CIR buffer in such a way that the reported first path index (\path{FP_INDEX}) in each CIR measurement is usually around 750 \cite{chorus}. The chipset also estimates \path{FP_INDEX} in each CIR measurement with a resolution of $\frac{1.0016 \text{ns}}{64}$, such that it is represented as a real number, having integer and fractional parts (see column \path{fp_index} in UWB datasets).
Now, since each CIR measurement basically has a different \path{FP_INDEX} value but the same sampling resolution of 1.0016 ns, the accumulated CIR measurements need to be aligned with their respective estimated \path{FP_INDEX}, and the latter can be shifted to be at the beginning of the CIR buffer, as shown in Fig. \ref{uwb_cir}. 
Furthermore, in order to remove outliers in the accumulated CIR, those CIR measurements where the number of accumulated preamble symbols (see column \path{rx_pream_count} in UWB datasets) is less than half of the number of transmitted preamble symbols (preamble length of 128 considered in the experiments) can be discarded \cite{uwbmultistatic2020}.

\subsection*{PWR}
Fig. \ref{signals_csi_uwb}(d)
illustrates the Doppler spectrogram collected by the 
PWR system from a given angle (see placement of "rx2", "rx3" and "rx4" in Fig. \ref{room_layout}). 
As mentioned previously, the PWR system uses the CSI transmitter (NUC3) as the signal source with an injection packet rate of 1600Hz. 
The coherent processing interval was set at 1 second, and sampling rate was set at 20MHz for each channel. 
The signatures in the Doppler spectrogram demonstrate the relative velocity between human and receiver antenna. 
The bi-static velocity is maximum in a monostatic layout (0 degree), and minimum in a forward scatter layout (180 degrees). 
We can see large differences between the signatures from the "sit/stand" and "walk" activities in terms of Doppler frequency shifts. 
This is because the velocity of the walking activity is much higher than other activities. The activities "lie down" and "stand from floor" have opposite Doppler signatures due to the opposing directions the human body undertakes when performing these activities. 
Such signatures can potentially be used for various machine learning applications such as activity recognition, people counting, localization, etc.

\subsection*{Kinect}
\begin{figure}[ht]
	\begin{center}
		\includegraphics[width=1  \textwidth]{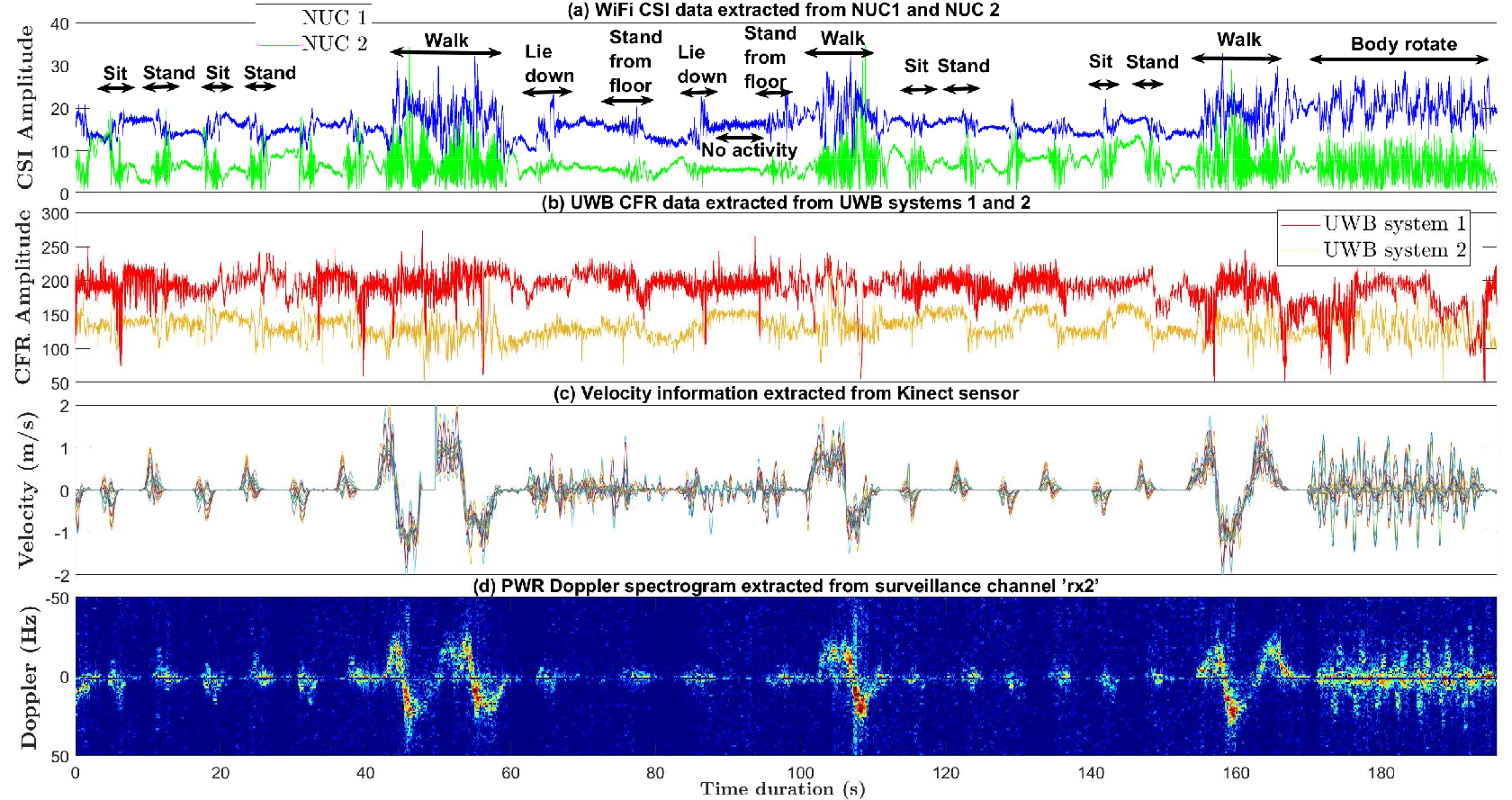} 
\caption{Signal analysis:  (a) WiFi CSI data (considering transmit antenna 1, receive antenna 1 and subcarrier 10), (b) UWB CFR data (considering the 10th CFR sample between nodes '0' and node '3' for UWB system 1 and nodes '1' and '2' for UWB system 2), (c) Velocity information extracted from Kinect sensor data and (d) PWR Doppler spectrogram extracted from surveillance channel 'rx2'.
Only a 196s portion of \textbf{exp018} is considered for the four synchronized modalities in this illustration.}

\label{signals_csi_uwb}
	\end{center}
\end{figure}

Fig. \ref{signals_csi_uwb}(c) illustrates the velocity-time profile of a human undergoing a set of activities. The motion profile is generated using the time-varying three-dimensional position information of different joints on the human body, such as the torso, arms, legs, and the head, at a frame rate of $10$ Hz. To mimic accurate radar reflections from the target, we assume the radar scattering centers lying approximately at the center of the bones joining these joints. Thus, the signatures in the velocity-time plot demonstrate the relative velocity between human scattering centers and the Kinect position.

Fig.\ref{signals_csi_uwb} (c)-(d), compare the velocity-time profile (generated using motion capture data) and the measured spectrograms, respectively, for a human undergoing a series of motions. The envelope of the velocity-time profile is visually very similar to the measured spectrogram indicating how well both systems capture the motion characteristics. For example, as the human sits down, we observe negative Doppler due to the bulk body motion. The positive micro-Doppler arises due to arm motion and legs moving slightly in the sensor direction while sitting down. After a 5sec delay, the human subject stands up from the chair, resulting in primarily positive Dopplers. Similarly, the latter part of the spectrogram presents signatures corresponding to a human transitioning from first walking to falling and then standing up from the ground to rotate his body while standing at a fixed position. 

In most realistic scenarios, the human motions might not be restricted to a single aspect angle with respect to the radar. In such scenarios, the spectrograms might differ significantly. It could be due to the shadowing of some part of the human body if captured at different angles. Therefore, we can leverage the animation data captured by the Kinect to feed as input to our human radar simulator, \textit{SimHumalator} and synthesize radar returns as a function of different target parameters (varying aspect angles) and different sensor parameters (varying bi-static angles and operational parameters such as waveform design and antenna characteristics). Such signatures can potentially augment otherwise limited real radar databases for various machine learning applications such as activity recognition, people counting, and identification. 


 \begin{figure}[t]
	\centering
	\subfigure[]
	{
		\includegraphics[width=65mm,scale=0.5]{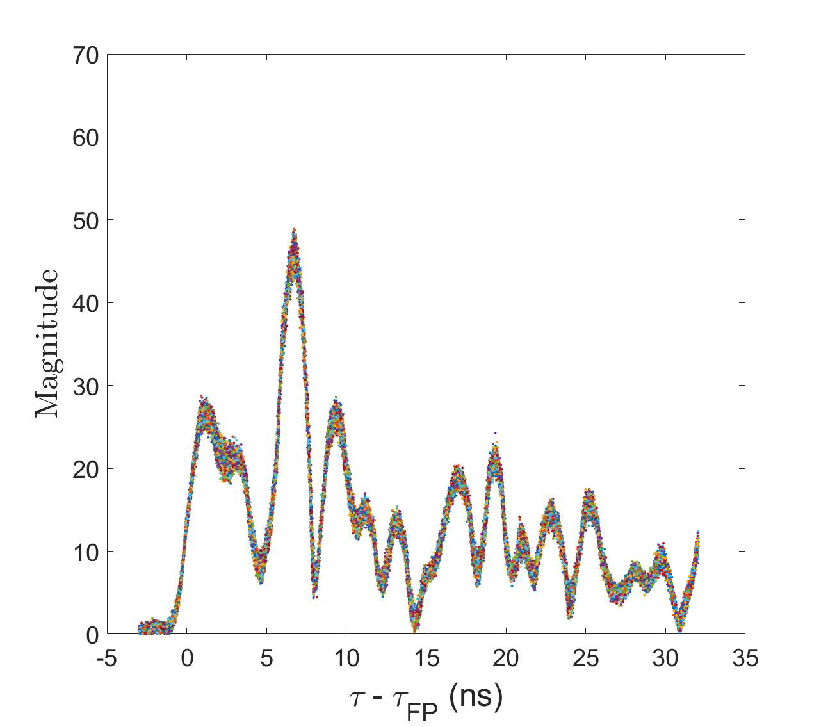}
		\label{uwb_cira}
	}	
\hspace{-2.1  \baselineskip}
	\subfigure[]
	{
		\includegraphics[width=65mm,scale=0.5]{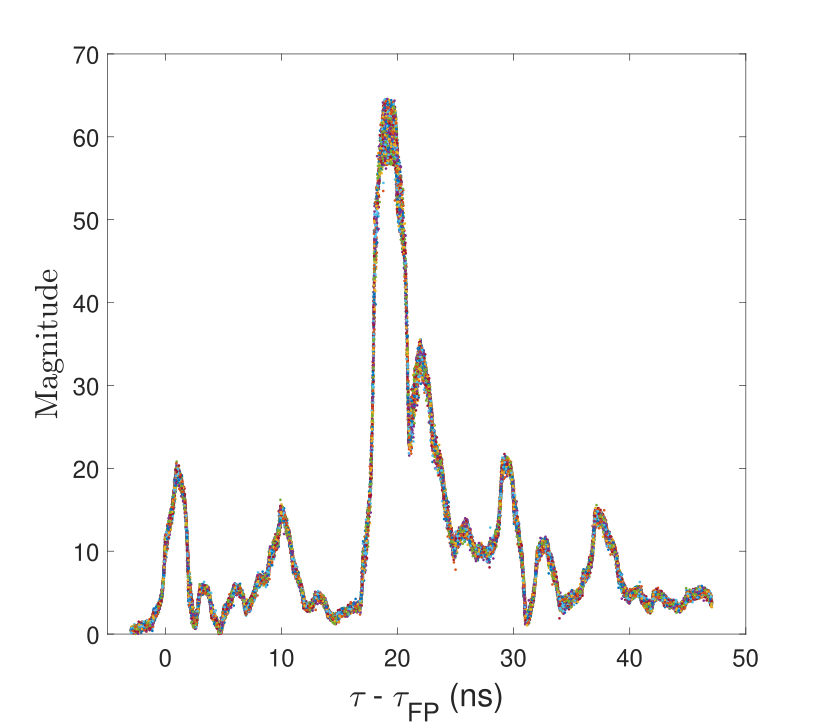}
		\label{uwb_cirb}
	}
\hspace{-2.1  \baselineskip}
	\subfigure[]
	{
		\includegraphics[width=65mm,scale=0.5]{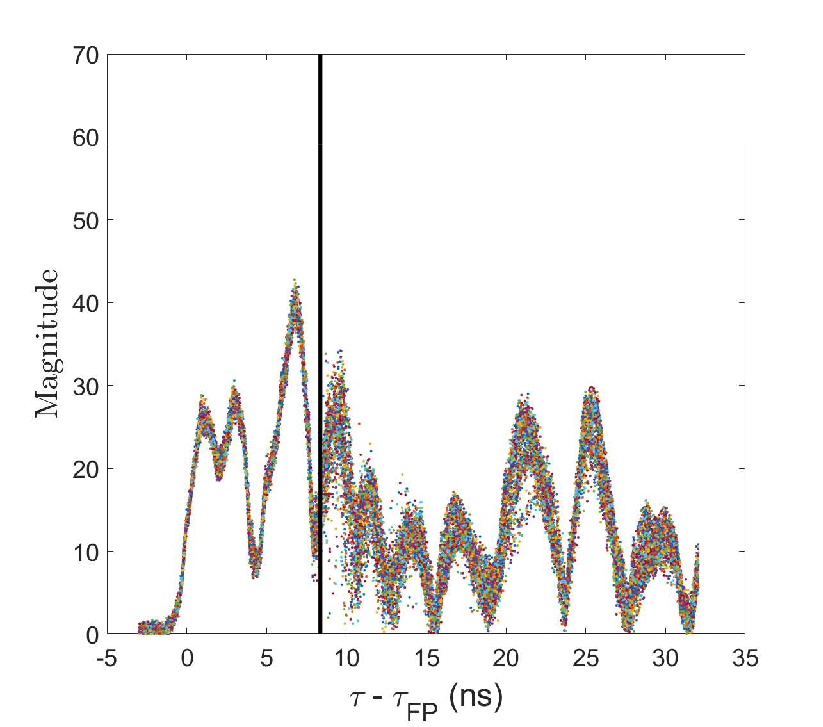}
		\label{uwb_circ}
	}
\hspace{-2.1  \baselineskip}
	\subfigure[]
	{
		\includegraphics[width=65mm,scale=0.5]{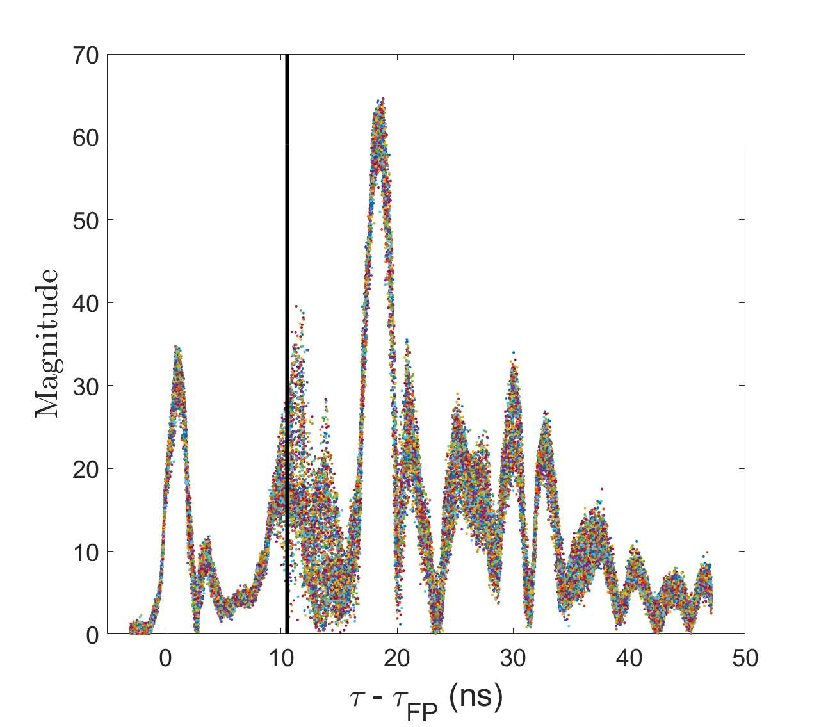}
		\label{uwb_cird}
	}
	\caption{1000 accumulated and aligned CIR  measurements in a 
	(a) static environment (\textbf{exp001}) recorded between nodes '1' and '3' of UWB system 1 (yellow nodes) 
	(b) static environment (\textbf{exp001}) recorded between nodes '2' and '3' of UWB system 2 (blue nodes)
	(c) dynamic environment (\textbf{exp003}) recorded between nodes '1' and '3' of UWB system 1 (yellow nodes) 
	(d) dynamic environment (\textbf{exp003}) recorded between nodes '2' and '3' of UWB system 2 (blue nodes). 
	Note: Bidirectional CIR data are reciprocal. $\tau_{FP}$ represents the first path (direct) signal time-of-flight between the pair of nodes.}
	\label{uwb_cir}
\end{figure}

 \begin{figure}[t]
	\centering
	\subfigure[]
	{
		\includegraphics[width=89mm,scale=0.5]{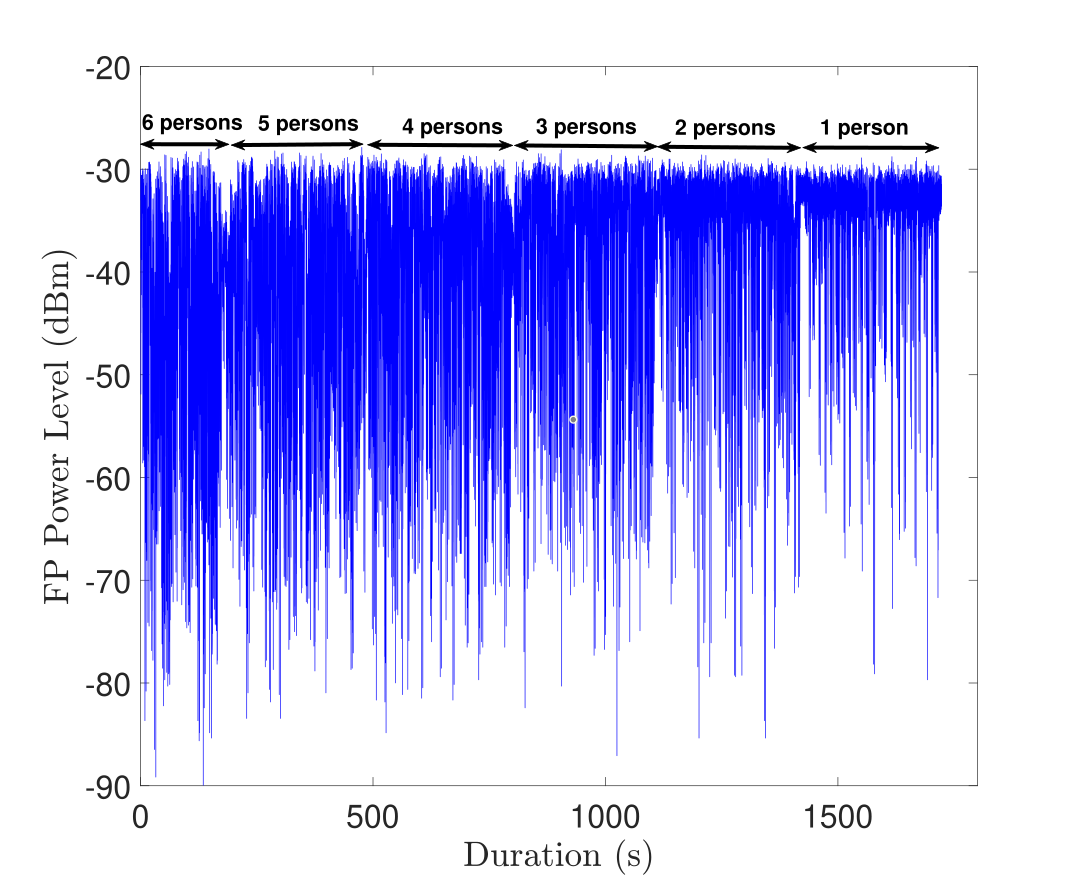}
		\label{fp_pow_figa}
	}	
\hspace{-2.1  \baselineskip}
	\subfigure[]
	{
		\includegraphics[width=89mm,scale=0.5]{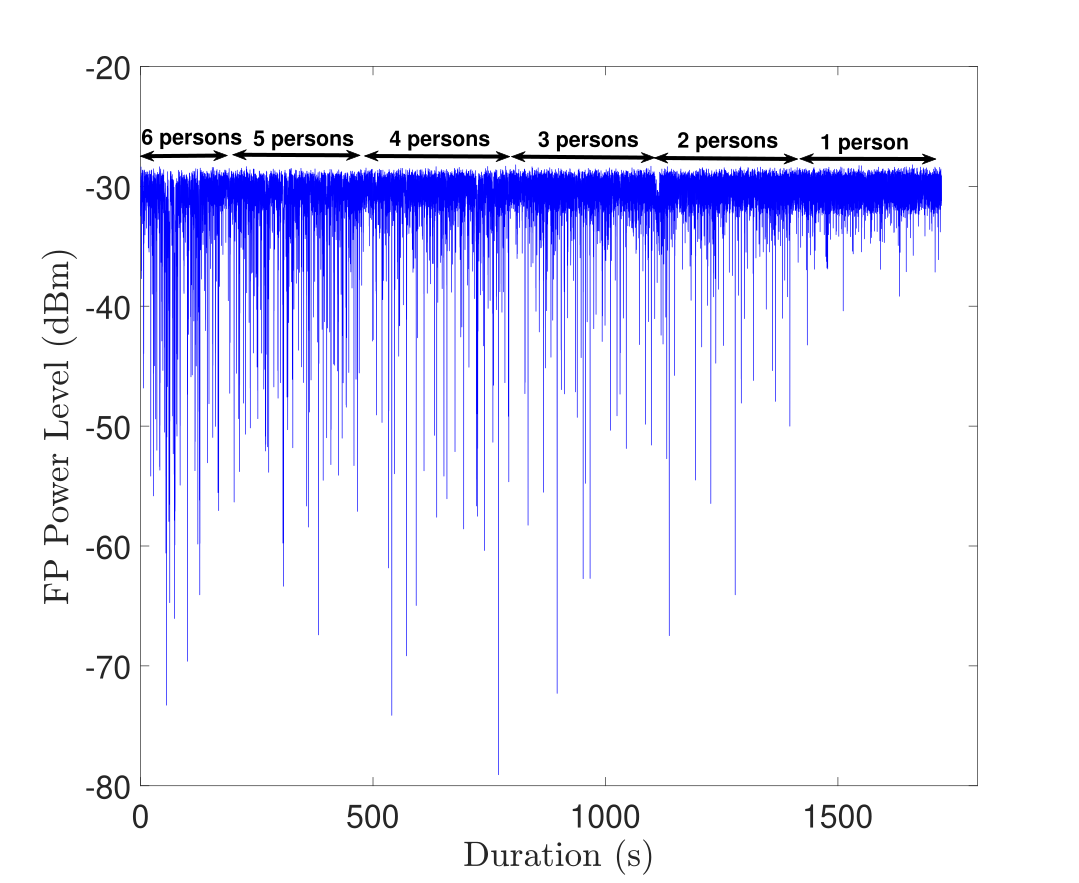}
		\label{fp_pow_figb}
	}
	\caption{First path power level (dBm) of UWB signal in crowd counting experiment (\textbf{exp028}) between nodes (a) '0' and '3' of UWB system 1 (yellow nodes) (b) nodes '3' and '4' of UWB system 2 (blue nodes).}
	\label{fp_pow_fig}
\end{figure}

\section*{Usage Notes}


The dataset repository is available 
at \url{https://figshare.com/s/c774748e127dcdecc667}.
The different directories are specified in
Table \ref{dataset_direc}.
Furthermore, the interested reader is encouraged to navigate to the \path{codes} directory
where
example scripts on how to load and analyze a specific modality data 
are included.
These are described in the following section.

\section*{Code availability}
Some Matlab and Python scripts have been made available in the \path{codes} directory for the users to replicate some of the figures in this Data Descriptor:

\begin{itemize}
\item  \path{plot_wificsi.m}: This script is used to load the complex WiFi CSI data recorded by each NUC device and visualize the amplitude variations over time, as illustrated in Fig. \ref{signals_csi_uwb}(a). The user can specify the start and stop timestamps and visualize the CSI stream in that time segment (for a given transmit antenna, receive antenna, and subcarrier index). Furthermore, for comparison purposes, the generated plots consist of the raw (unfiltered) CSI data and those which have been denoised using DWT.

\item  \path{plot_uwb.m}: This script is used to load the complex CIR  data recorded by each passive UWB system, convert it into CFR using FFT and visualize the amplitude variations over a given time segment (between a given pair of UWB nodes), as illustrated in Fig. \ref{signals_csi_uwb}(b). Furthermore, this script allows the users to plot the aligned CIR measurements as shown in Fig. \ref{uwb_cir}.

\item  \path{plot_uwb_fppow_crowdcount.m}: This script is used to load the UWB data for the crowd counting experiment (\path{exp028}) and plot the first path power level (in dBm) over time for each UWB system (between a given pair of UWB nodes), as illustrated in Fig. \ref{fp_pow_fig}.

\item  \path{plot_PWR_demonstration.m} and \path{plot_pwr_spectrogram.py} : These scripts allow the users to visualize the PWR spectrograms from the three surveillance channels: "rx2" (as illustrated in Fig. \ref{signals_csi_uwb}(d).), "rx3" and "rx4", as a function of time and Doppler.

\item  \path{plot_kinect_data.m}: This script allows the user to plot the motion capture data (as a function of velocity versus time) from one of the two Kinect systems, as illustrated in Fig. \ref{signals_csi_uwb}(c). Furthermore, the users can visualize the stick (skeletal) representation of the kinect motion capture data as an animation over the specified time segment.

\item  \path{oddet.py}: This python script allows the user
to extract only the modalities and features needed rather than needing to take the entire files and then stripping out unused features. With this python script, one can select the modality, experiment number and features needed through the command line interface. Additionally if a specific set of features are required, one can also specify all the columns needed through YAML configurations which will allow the user to curate the dataset to the format that more closely suits the usage. This python script is available at the following GitHub repository: \url{https://github.com/RogetK/ODDET}.
\end{itemize}


\bibliography{main.bib}

\begin{thebibliography}{10}
\urlstyle{rm}
\expandafter\ifx\csname url\endcsname\relax
  \def\url#1{\texttt{#1}}\fi
\expandafter\ifx\csname urlprefix\endcsname\relax\def\urlprefix{URL }\fi
\expandafter\ifx\csname doiprefix\endcsname\relax\def\doiprefix{DOI: }\fi
\providecommand{\bibinfo}[2]{#2}
\providecommand{\eprint}[2][]{\url{#2}}

\bibitem{accgyr}
\bibinfo{author}{Zhao, Y.}, \bibinfo{author}{Yang, R.},
  \bibinfo{author}{Chevalier, G.} \& \bibinfo{author}{Gong, M.}
\newblock \bibinfo{journal}{\bibinfo{title}{Deep residual bidir-lstm for human
  activity recognition using wearable sensors}}.
\newblock {\emph{\JournalTitle{CoRR}}}
  \textbf{\bibinfo{volume}{abs/1708.08989}} (\bibinfo{year}{2017}).
\newblock \eprint{1708.08989}.

\bibitem{kinect2018}
\bibinfo{author}{Gavrilova, M.~L.}, \bibinfo{author}{Wang, Y.},
  \bibinfo{author}{Ahmed, F.} \& \bibinfo{author}{Polash~Paul, P.}
\newblock \bibinfo{journal}{\bibinfo{title}{Kinect sensor gesture and activity
  recognition: New applications for consumer cognitive systems}}.
\newblock {\emph{\JournalTitle{IEEE Consumer Electronics Magazine}}}
  \textbf{\bibinfo{volume}{7}}, \bibinfo{pages}{88--94},
  \url{10.1109/MCE.2017.2755498} (\bibinfo{year}{2018}).

\bibitem{translationresilientcsi}
\bibinfo{author}{Bocus, M.~J.} \emph{et~al.}
\newblock \bibinfo{title}{Translation resilient opportunistic wifi sensing}.
\newblock In \emph{\bibinfo{booktitle}{2020 25th International Conference on
  Pattern Recognition (ICPR)}}, \bibinfo{pages}{5627--5633},
  \url{10.1109/ICPR48806.2021.9412263} (\bibinfo{year}{2021}).

\bibitem{taxonomycsi}
\bibinfo{author}{Li, W.} \emph{et~al.}
\newblock \bibinfo{title}{A taxonomy of wifi sensing: Csi vs passive wifi
  radar}.
\newblock In \emph{\bibinfo{booktitle}{2020 IEEE Globecom Workshops (GC
  Wkshps}}, \bibinfo{pages}{1--6}, \url{10.1109/GCWkshps50303.2020.9367546}
  (\bibinfo{year}{2020}).

\bibitem{bocus2021uwb}
\bibinfo{author}{Bocus, M.~J.}, \bibinfo{author}{Chetty, K.} \&
  \bibinfo{author}{Piechocki, R.~J.}
\newblock \bibinfo{title}{Uwb and wifi systems as passive opportunistic
  activity sensing radars}.
\newblock In \emph{\bibinfo{booktitle}{2021 IEEE Radar Conference
  (RadarConf21)}}, \bibinfo{pages}{1--6} (\bibinfo{organization}{IEEE},
  \bibinfo{year}{2021}).

\bibitem{ALAZRAI}
\bibinfo{author}{Alazrai, R.}, \bibinfo{author}{Awad, A.},
  \bibinfo{author}{Alsaify, B.}, \bibinfo{author}{Hababeh, M.} \&
  \bibinfo{author}{Daoud, M.~I.}
\newblock \bibinfo{journal}{\bibinfo{title}{A dataset for wi-fi-based
  human-to-human interaction recognition}}.
\newblock {\emph{\JournalTitle{Data in Brief}}} \textbf{\bibinfo{volume}{31}},
  \bibinfo{pages}{105668}, \url{https://doi.org/10.1016/j.dib.2020.105668}
  (\bibinfo{year}{2020}).

\bibitem{Alsaify2020}
\bibinfo{author}{Alsaify, B.~A.}, \bibinfo{author}{Almazari, M.~M.},
  \bibinfo{author}{Alazrai, R.} \& \bibinfo{author}{Daoud, M.~I.}
\newblock \bibinfo{journal}{\bibinfo{title}{A dataset for wi-fi-based human
  activity recognition in line-of-sight and non-line-of-sight indoor
  environments}}.
\newblock {\emph{\JournalTitle{Data in Brief}}} \textbf{\bibinfo{volume}{33}},
  \bibinfo{pages}{106534}, \url{https://doi.org/10.1016/j.dib.2020.106534}
  (\bibinfo{year}{2020}).

\bibitem{signfi}
\bibinfo{author}{Ma, Y.}, \bibinfo{author}{Zhou, G.}, \bibinfo{author}{Wang,
  S.}, \bibinfo{author}{Zhao, H.} \& \bibinfo{author}{Jung, W.}
\newblock \bibinfo{journal}{\bibinfo{title}{Signfi: Sign language recognition
  using wifi}}.
\newblock {\emph{\JournalTitle{Proc. ACM Interact. Mob. Wearable Ubiquitous
  Technol.}}} \textbf{\bibinfo{volume}{2}}, \url{10.1145/3191755}
  (\bibinfo{year}{2018}).

\bibitem{FallDeFi}
\bibinfo{author}{Palipana, S.}, \bibinfo{author}{Rojas, D.},
  \bibinfo{author}{Agrawal, P.} \& \bibinfo{author}{Pesch, D.}
\newblock \bibinfo{journal}{\bibinfo{title}{Falldefi: Ubiquitous fall detection
  using commodity wi-fi devices}}.
\newblock {\emph{\JournalTitle{Proc. ACM Interact. Mob. Wearable Ubiquitous
  Technol.}}} \textbf{\bibinfo{volume}{1}}, \url{10.1145/3161183}
  (\bibinfo{year}{2018}).

\bibitem{spotfi}
\bibinfo{author}{Kotaru, M.}, \bibinfo{author}{Joshi, K.},
  \bibinfo{author}{Bharadia, D.} \& \bibinfo{author}{Katti, S.}
\newblock \bibinfo{journal}{\bibinfo{title}{Spotfi: Decimeter level
  localization using wifi}}.
\newblock {\emph{\JournalTitle{SIGCOMM Comput. Commun. Rev.}}}
  \textbf{\bibinfo{volume}{45}}, \bibinfo{pages}{269–282},
  \url{10.1145/2829988.2787487} (\bibinfo{year}{2015}).

\bibitem{UWB-gestures}
\bibinfo{author}{Ahmed, S.}, \bibinfo{author}{Wang, D.}, \bibinfo{author}{Park,
  J.} \& \bibinfo{author}{Cho, S.~H.}
\newblock \bibinfo{journal}{\bibinfo{title}{Uwb-gestures, a public dataset of
  dynamic hand gestures acquired using impulse radar sensors}}.
\newblock {\emph{\JournalTitle{Scientific Data}}} \textbf{\bibinfo{volume}{8}},
  \bibinfo{pages}{102} (\bibinfo{year}{2021}).

\bibitem{klemenbregar5}
\bibinfo{author}{Bregar, K.}, \bibinfo{author}{Hrovat, A.} \&
  \bibinfo{author}{Mohorčič, M.}
\newblock \bibinfo{title}{Uwb motion detection data set},
  \url{10.5281/zenodo.4613125} (\bibinfo{year}{2021}).

\bibitem{uwbmultistatic2020}
\bibinfo{author}{Ledergerber, A.} \& \bibinfo{author}{D’Andrea, R.}
\newblock \bibinfo{journal}{\bibinfo{title}{A multi-static radar network with
  ultra-wideband radio-equipped devices}}.
\newblock {\emph{\JournalTitle{Sensors}}} \textbf{\bibinfo{volume}{20}},
  \url{10.3390/s20061599} (\bibinfo{year}{2020}).

\bibitem{uwbpeoplecount}
\bibinfo{author}{Yang, X.}, \bibinfo{author}{Yin, W.}, \bibinfo{author}{Li, L.}
  \& \bibinfo{author}{Zhang, L.}
\newblock \bibinfo{journal}{\bibinfo{title}{Dense people counting using ir-uwb
  radar with a hybrid feature extraction method}}.
\newblock {\emph{\JournalTitle{IEEE Geoscience and Remote Sensing Letters}}}
  \textbf{\bibinfo{volume}{16}}, \bibinfo{pages}{30--34},
  \url{10.1109/LGRS.2018.2869287} (\bibinfo{year}{2019}).

\bibitem{Byrne2018}
\bibinfo{author}{Byrne, D.}, \bibinfo{author}{Kozlowski, M.},
  \bibinfo{author}{Santos-Rodriguez, R.}, \bibinfo{author}{Piechocki, R.} \&
  \bibinfo{author}{Craddock, I.}
\newblock \bibinfo{journal}{\bibinfo{title}{Residential wearable rssi and
  accelerometer measurements with detailed location annotations}}.
\newblock {\emph{\JournalTitle{Scientific Data}}} \textbf{\bibinfo{volume}{5}},
  \bibinfo{pages}{180168} (\bibinfo{year}{2018}).

\bibitem{GarciaGonzalez2020}
\bibinfo{author}{Garcia-Gonzalez, D.}, \bibinfo{author}{Rivero, D.},
  \bibinfo{author}{Fernandez-Blanco, E.} \& \bibinfo{author}{Luaces, M.~R.}
\newblock \bibinfo{journal}{\bibinfo{title}{A public domain dataset for
  real-life human activity recognition using smartphone sensors}}.
\newblock {\emph{\JournalTitle{Sensors}}} \textbf{\bibinfo{volume}{20}},
  \url{10.3390/s20082200} (\bibinfo{year}{2020}).

\bibitem{kinectdataset}
\bibinfo{author}{Chen, C.}, \bibinfo{author}{Jafari, R.} \&
  \bibinfo{author}{Kehtarnavaz, N.}
\newblock \bibinfo{title}{Utd-mhad: A multimodal dataset for human action
  recognition utilizing a depth camera and a wearable inertial sensor}.
\newblock In \emph{\bibinfo{booktitle}{2015 IEEE International Conference on
  Image Processing (ICIP)}}, \bibinfo{pages}{168--172},
  \url{10.1109/ICIP.2015.7350781} (\bibinfo{year}{2015}).

\bibitem{vishwakarma2021simhumalator}
\bibinfo{author}{Vishwakarma, S.} \emph{et~al.}
\newblock \bibinfo{journal}{\bibinfo{title}{Simhumalator: An open source wifi
  based passive radar human simulator for activity recognition}}.
\newblock {\emph{\JournalTitle{arXiv preprint arXiv:2103.01677}}}
  (\bibinfo{year}{2021}).

\bibitem{csitool}
\bibinfo{author}{Halperin, D.}, \bibinfo{author}{Hu, W.},
  \bibinfo{author}{Sheth, A.} \& \bibinfo{author}{Wetherall, D.}
\newblock \bibinfo{journal}{\bibinfo{title}{Tool release: Gathering 802.11n
  traces with channel state information}}.
\newblock {\emph{\JournalTitle{SIGCOMM Comput. Commun. Rev.}}}
  \textbf{\bibinfo{volume}{41}}, \bibinfo{pages}{53},
  \url{10.1145/1925861.1925870} (\bibinfo{year}{2011}).

\bibitem{Linux80228}
\bibinfo{author}{Halperin, D.}
\newblock \bibinfo{title}{Linux 802.11n csi tool}.
\newblock
  \bibinfo{howpublished}{\url{https://dhalperi.github.io/linux-80211n-csitool/}}.
\newblock \bibinfo{note}{(Accessed on 07/02/2021)}.

\bibitem{EVK1000}
\bibinfo{title}{Evk1000 evaluation kit - decawave}.
\newblock
  \bibinfo{howpublished}{\url{https://www.decawave.com/product/evk1000-evaluation-kit/}}.
\newblock \bibinfo{note}{(Accessed on 07/02/2021)}.

\bibitem{MDEK100157}
\bibinfo{author}{Decawave}.
\newblock \bibinfo{title}{Mdek1001 development kit - decawave}.
\newblock
  \bibinfo{howpublished}{\url{https://www.decawave.com/product/mdek1001-deployment-kit/}}.
\newblock \bibinfo{note}{(Accessed on 07/02/2021)}.

\bibitem{DW1000ma}
\bibinfo{title}{Dw1000 user manual - decawave}.
\newblock
  \bibinfo{howpublished}{\url{https://www.decawave.com/dw1000/usermanual/}}.
\newblock \bibinfo{note}{(Accessed on 01/07/2021)}.

\bibitem{safetycir2016}
\bibinfo{author}{{Moschevikin}, A.}, \bibinfo{author}{{Tsvetkov}, E.},
  \bibinfo{author}{{Alekseev}, A.} \& \bibinfo{author}{{Sikora}, A.}
\newblock \bibinfo{title}{Investigations on passive channel impulse response of
  ultra wide band signals for monitoring and safety applications}.
\newblock In \emph{\bibinfo{booktitle}{2016 3rd International Symposium on
  Wireless Systems within the Conferences on Intelligent Data Acquisition and
  Advanced Computing Systems (IDAACS-SWS)}}, \bibinfo{pages}{97--104},
  \url{10.1109/IDAACS-SWS.2016.7805795} (\bibinfo{year}{2016}).

\bibitem{USRP2945}
\bibinfo{title}{Usrp-2945}.
\newblock
  \bibinfo{howpublished}{https://www.ni.com/en-gb/support/model.usrp-2945.html}.
\newblock \bibinfo{note}{(Accessed on 07/05/2021)}.

\bibitem{li2020passive}
\bibinfo{author}{Li, W.}, \bibinfo{author}{Piechocki, R.~J.},
  \bibinfo{author}{Woodbridge, K.}, \bibinfo{author}{Tang, C.} \&
  \bibinfo{author}{Chetty, K.}
\newblock \bibinfo{journal}{\bibinfo{title}{Passive wifi radar for human
  sensing using a stand-alone access point}}.
\newblock {\emph{\JournalTitle{IEEE Transactions on Geoscience and Remote
  Sensing}}} \textbf{\bibinfo{volume}{59}}, \bibinfo{pages}{1986--1998}
  (\bibinfo{year}{2020}).

\bibitem{Kinect}
\bibinfo{author}{Microsoft, X.}
\newblock \bibinfo{title}{Kinect sensor} (\bibinfo{year}{2021}).

\bibitem{chai2012general}
\bibinfo{author}{Chai, Y.}
\newblock \emph{\bibinfo{title}{A General Framework for Motion Sensor Based Web
  Services}}.
\newblock Ph.D. thesis, \bibinfo{school}{University of Saskatchewan}
  (\bibinfo{year}{2012}).

\bibitem{kramer2012hacking}
\bibinfo{author}{Kramer, J.}, \bibinfo{author}{Burrus, N.},
  \bibinfo{author}{Echtler, F.}, \bibinfo{author}{Daniel, H.~C.} \&
  \bibinfo{author}{Parker, M.}
\newblock \emph{\bibinfo{title}{Hacking the kinect}}, vol.
  \bibinfo{volume}{268} (\bibinfo{publisher}{Springer}, \bibinfo{year}{2012}).

\bibitem{nyberg2014development}
\bibinfo{author}{Nyberg, R.}
\newblock \bibinfo{title}{Development of a mobile robot platform}
  (\bibinfo{year}{2014}).

\bibitem{tang2021augmenting}
\bibinfo{author}{Tang, C.} \emph{et~al.}
\newblock \bibinfo{title}{Augmenting experimental data with simulations to
  improve activity classification in healthcare monitoring}.
\newblock In \emph{\bibinfo{booktitle}{2021 IEEE Radar Conference
  (RadarConf21)}}, \bibinfo{pages}{1--6} (\bibinfo{organization}{IEEE},
  \bibinfo{year}{2021}).

\bibitem{vishwakarma2021gan}
\bibinfo{author}{Vishwakarma, S.} \emph{et~al.}
\newblock \bibinfo{title}{Gan based noise generation to aid activity
  recognition when augmenting measured wifi radar data with simulations}.
\newblock In \emph{\bibinfo{booktitle}{IEEE International Conference on
  Communications}} (\bibinfo{year}{2021}).

\bibitem{lau2021selfsupervised}
\bibinfo{author}{Lau, H.-S.}, \bibinfo{author}{McConville, R.},
  \bibinfo{author}{Bocus, M.~J.}, \bibinfo{author}{Piechocki, R.~J.} \&
  \bibinfo{author}{Santos-Rodriguez, R.}
\newblock \bibinfo{title}{Self-supervised wifi-based activity recognition}
  (\bibinfo{year}{2021}).
\newblock \eprint{2104.09072}.

\bibitem{chorus}
\bibinfo{author}{{Corbalán}, P.}, \bibinfo{author}{{Picco}, G.~P.} \&
  \bibinfo{author}{{Palipana}, S.}
\newblock \bibinfo{title}{{Chorus}: {UWB} concurrent transmissions for
  {GPS}-like passive localization of countless targets}.
\newblock In \emph{\bibinfo{booktitle}{2019 18th ACM/IEEE International
  Conference on Information Processing in Sensor Networks (IPSN)}},
  \bibinfo{pages}{133--144}, \url{10.1145/3302506.3310395}
  (\bibinfo{year}{2019}).

\end{thebibliography}

\section*{Acknowledgements}  
This work was funded under the OPERA Project, the UK Engineering and Physical Sciences Research Council (EPSRC), Grant EP/R018677/1. 

\section*{Author contributions statement}
M.B. and W.L. conceived and planned the experiment, R.P. and K.C. reviewed the experiment plan and supervised the whole experiment. 
M.B. and R.P. made the necessary applications and obtained the risk assessment and ethics approvals prior to the experiment.
M.B., W.L., S.V., R.K., T.C., K.C. and R.P. 
conducted the experiment and recorded
the experimental data from the different modalities.
M.B. and S.V. developed human activity annotation tools on Matlab. 
M.B., W.L. and S.V. curated and analyzed the datasets. 
M.B., W.L., T.C., R.K. and S.V. wrote the scripts in Matlab and Python for loading and analyzing the dataset from each modality.
M.B. took the lead in preparing the Data Descriptor and was assisted by W.L. and S.V. for populating the various sections in this manuscript. 
All authors reviewed the manuscript, provided constructive feedback and assisted with the editing of the manuscript prior to submission.

\section*{Competing interests}
The authors declare no competing interests.

\end{document}